%Paper: cond-mat/9408059
%From: "Deniz Ertas" <deniz@cmt8.mit.edu>
%Date: Fri, 19 Aug 1994 11:19:03 -0400

%%
%%  To compile properly, save the portion of the file after \end
%%  as figures.uu and execute the shell to obtain epsf.sty and
%%  the PostScript figures.
%%
%%	Proceedings for the Newton Institute 94 %%%%%%%%
\input epsf
%-----------------Plenum Template----------------------------------------%
\magnification=\magstep1
%\nopagenumbers
\scrollmode
\overfullrule=0mm
\hsize=6.1truein \vsize=9.4truein \hoffset=.2truein
%-----------------Equation numbers----------------------------------------%
\global\newcount\meqno \global\meqno=1
\def\eqn#1#2{\xdef #1{(\the\meqno)}\global\advance\meqno by1 $$#2\eqno#1$$}
%-----------------Reference numbers---------------------------------------%
\global\newcount\refno \global\refno=1 \newwrite\rfile
\def\ref#1#2{$^{\the\refno}$\nref#1{#2}}%
\def\nref#1#2{\xdef#1{$^{\the\refno}$}%
\ifnum\refno=1\immediate\openout\rfile=refs.tmp\fi%
\immediate\write\rfile{\noexpand\item{\the\refno .\ }#2}%
\global\advance\refno by1}

\def\rsc{$^,$} 
%---Figure numbers----------------------------------------%
\global\newcount\figno \global\figno=1
\def\figure#1#2#3{\midinsert\centerline{\epsffile{#2}}
\xdef #1{\the\figno}\global\advance\figno by1
\smallskip\centerline{\vtop{\hsize 4.1in \noindent {\bf Figure #1.}~#3}}
\bigskip\endinsert}
\def\nofigure#1#2#3{\midinsert\vskip#2 truecm
\xdef #1{\chapsym.\the\figno}\global\advance\figno by1
\hfil{{\bf #1}~#3}\hfil\endinsert}
%---Bold math symbols-------------------------------------%
\font\bms=cmbsy10
%%%%%%%%%%%%%%%%%%%%%%%%%
\def\frac#1#2{{#1\over #2}}

\def\bx{{\bf x}}
\def\bq{{\bf q}}
\def\br{{\bf r}}

\def\bF{{\bf F}}

\def\bv{{\bf v}}

\def\bfr{{\bf f}}
\def\rh{{\hat r}}
\def\Rh{{\hat R}}

\def\bR{{\bf R}}
\def\bRh{{\bf \hat R}}
% These don't work!!
%\def\beps{{\hbox{\boldmath$\varepsilon$}}}
%\def\bchi{{\bms \chi}}
\def\beps{{\bms \varepsilon}}

\def\dx{{d^d\bx\,}}
\def\dt{{dt\,}}
\def\dq{{d^d\bq \over (2\pi)^d}}
\def\dw{{d\omega \over 2\pi}}
\def\w{{\omega}}
\def\K{{\cal K}}
%-----------------Other definitions---------------------------------------
\def\del{\partial}    \def\frac#1#2{{#1\over#2}}
   \def\br{{\bf r}}
\def\para{\parallel}
\def\hl{{r_\parallel}}
\def\htr{{r_\perp}}
\def\lal{\lambda_\para} \def\lat{\lambda_\perp} \def\lalt{\lambda_\times}
\def\Kl{{K_\para}} \def\Kt{{K_\perp}}
\def\fl{{f_\para}} \def\ft{{f_\perp}}
\def\Tl{{T_\para}} \def\Tt{{T_\perp}}
\def\zetal{{\zeta_\para}} \def\zetat{{\zeta_\perp}}
\def\zl{{z_\para}} \def\zt{{z_\perp}}
%-------------------------------------------------------------------------%
%%
%-----------------TITLE PAGE----------------------------------------------%
\parindent 0pt
%\nopagenumbers
{\bf NONEQUILIBRIUM DYNAMICS OF FLUCTUATING LINES}
%\vskip .15truein
%{\bf FLUCTUATING LINES}
\vskip .30truein \hskip 1.15truein
Mehran Kardar and Deniz Erta\c s\hfill
\vskip .15truein \vskip -.5truecm \null \hskip 1.15truein
Department of Physics \hfil\break \null \hskip 1.15truein
Massachusetts Institute of Technology \hfil\break \null \hskip 1.15truein
Cambridge, Massachusetts 02139, USA \hfil\break
\vskip .15truein
{\bf ABSTRACT} \smallskip
\parindent 20pt
%\nopagenumbers
{These notes are for the proceedings of the NATO Advanced Study
Insitutue on {\it Scale Invariance, Interfaces, and Non--Equilibrium Dynamics},
held at the Isaac Newton Institute, 20--30 June 1994.
The four lectures  address a number of issues related to dynamic
fluctuations of lines in non-equilibrium circumstances. The
first two are devoted to the critical behavior of contact lines
and flux lines depinning from impurities. It is emphasized that
anisotropies in the medium lead to different universality classes.
The importance of nonlinearities for moving lines are discussed
in the context of flux lines and polymers in the last two lectures.
A dynamic form birefringence is predicted for drifting polymers.}
%KEYWORDS: non-equilibrium dynamics / critical phenomena /
%pinning / depinning / interfaces / domain walls / contact line /
%flux line / contact angle / dynamic roughening / renormalization
%group / anisotropy / polymers / birefringence
%----------------Main text------------------------------------------------%
%\nopagenumbers
\vskip .15truein
\noindent{\bf I. DEPINNING OF A LINE IN TWO DIMENSIONS}\smallskip\par
Depinning is a non-equilibrium critical phenomenon involving an external force
and a pinning potential. When the force is weak the system is stationary,
trapped in a metastable state. Beyond a threshold force the (last) metastable
state disappears and the system starts to move. While there are many
macroscopic
mechanical examples, our interest stems from condensed matter systems
such as Charge Density Waves (CDWs)\ref
\rCDWs{H. Fukuyama and P. A. Lee, Phys. Rev. B {\bf 17}, 535 (1978);
P. A. Lee and T. M. Rice, Phys. Rev. B {\bf 19}, 3970 (1979).},
interfaces\ref
\rIntdepin{R. Bruinsma and G. Aeppli, Phys. Rev. Lett. {\bf 52}, 1547 (1984);
J. Koplik and H. Levine, Phys. Rev. B {\bf 32}, 280 (1985).},
and contact lines\ref
\rDeGennes{P.G.~de~Gennes, Rev. Mod. Phys. {\bf 57},
827 (1985).}.
In CDWs, the controlling parameter is the external voltage. A finite CDW
current
appears only beyond a threshold applied voltage. Interfaces in porous
media, domain walls in random magnets, are stationary unless the applied
force (magnetic field) is sufficiently strong. A key feature of these
examples is that they involve the {\it collective} depinning of many degrees
of freedom that are elastically coupled. As such these problems belong to
the realm of collective critical phenomena, characterized by universal
scaling laws. We shall introduce these laws and the corresponding
exponents below for the depinning of a line (interface or contact line).

\epsfysize=5cm\figure\CL{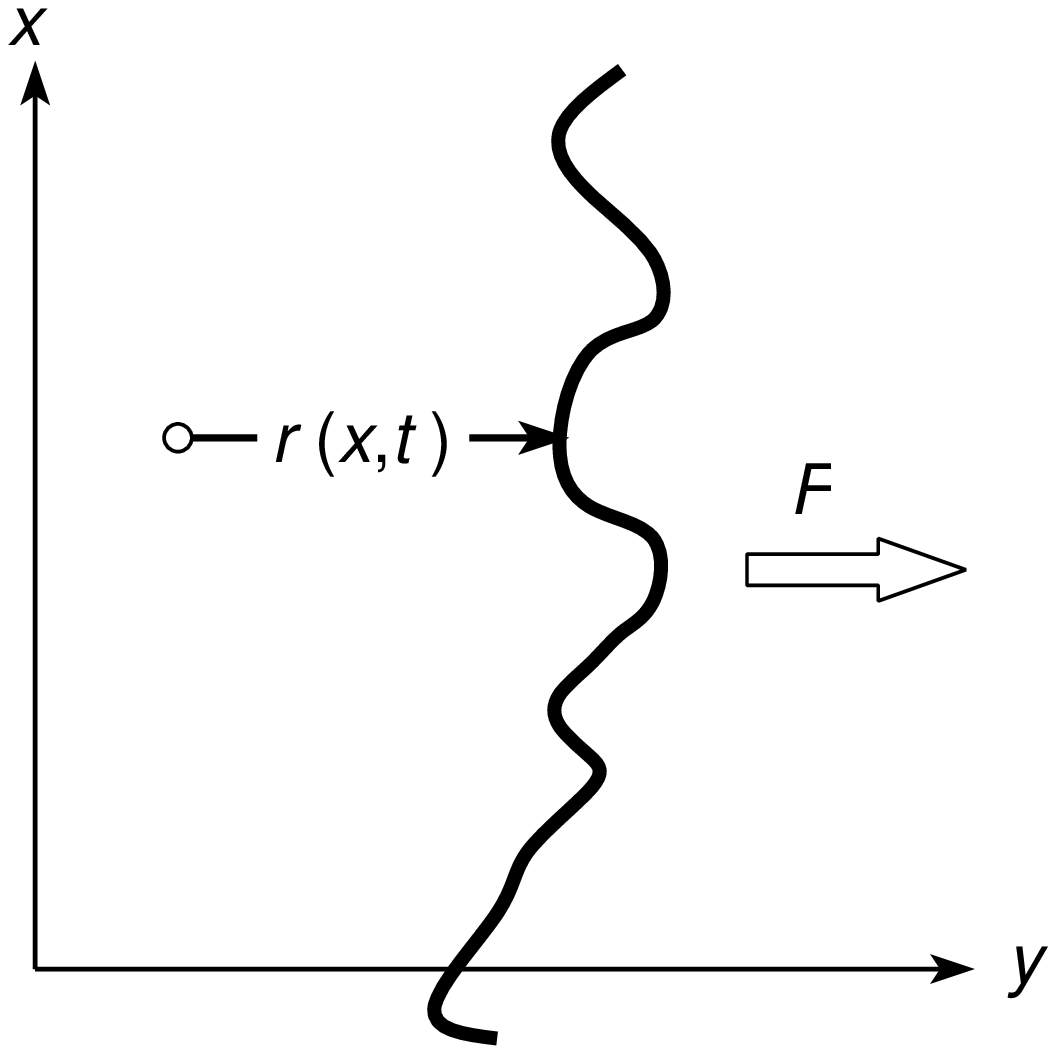}{Geometry of the line in two dimensions.}
Consider a line in two dimensions, oriented along the $x$ direction, and
fluctuating along a perpendicular $y$ direction. The configuration of the
line at time $t$ is described by the function $r(x,t)$.
The function $r$  is assumed
to be single valued, thus excluding configurations with overhangs.  In
many cases\rIntdepin, the evolution of the curve satisfies an equation of the
form
\eqn\eGen{{d{r(x,t)} \over dt}=F+f(x,r)+{\cal K}[r].}
The first term is a uniform applied force which is also the external control
parameter. Fluctuations in the force due to randomness and impurities
are represented by the second term. With the assumption that the medium
is on average translationally invariant, the average of $f$ can be set to zero.
The final term describes the elastic forces between different parts of the
line. Short range interactions can be described by a gradient expansion;
for example, a line tension leads to ${\cal K}[r(x)]=\nabla^2 r$ or
${\cal K}[r(q)]=-q^2 r(q)$ for the Fourier modes. The surface of a drop
of non--wetting liquid terminates at a {\it contact line} on a solid
substrate\rDeGennes.
Deformations of the contact line are accompanied by distortions of the
liquid/gas surface. As shown by Joanny and de Gennes\ref
\rJoanny{J.F.~Joanny and P.G.~de~Gennes, J. Chem. Phys.
{\bf 81}, 552 (1984).},
the resulting energy and forces are {\it non--local}, described by ${\cal
K}[r(q)]=-|q| r(q)$.
More generally we shall consider the linear operator
${\cal K}[r(q)]=-|q|^\sigma r(q)$,
which interpolates between the above two cases as $\sigma$ changes
from one to two.

\epsfysize=5cm\figure\fVvsF{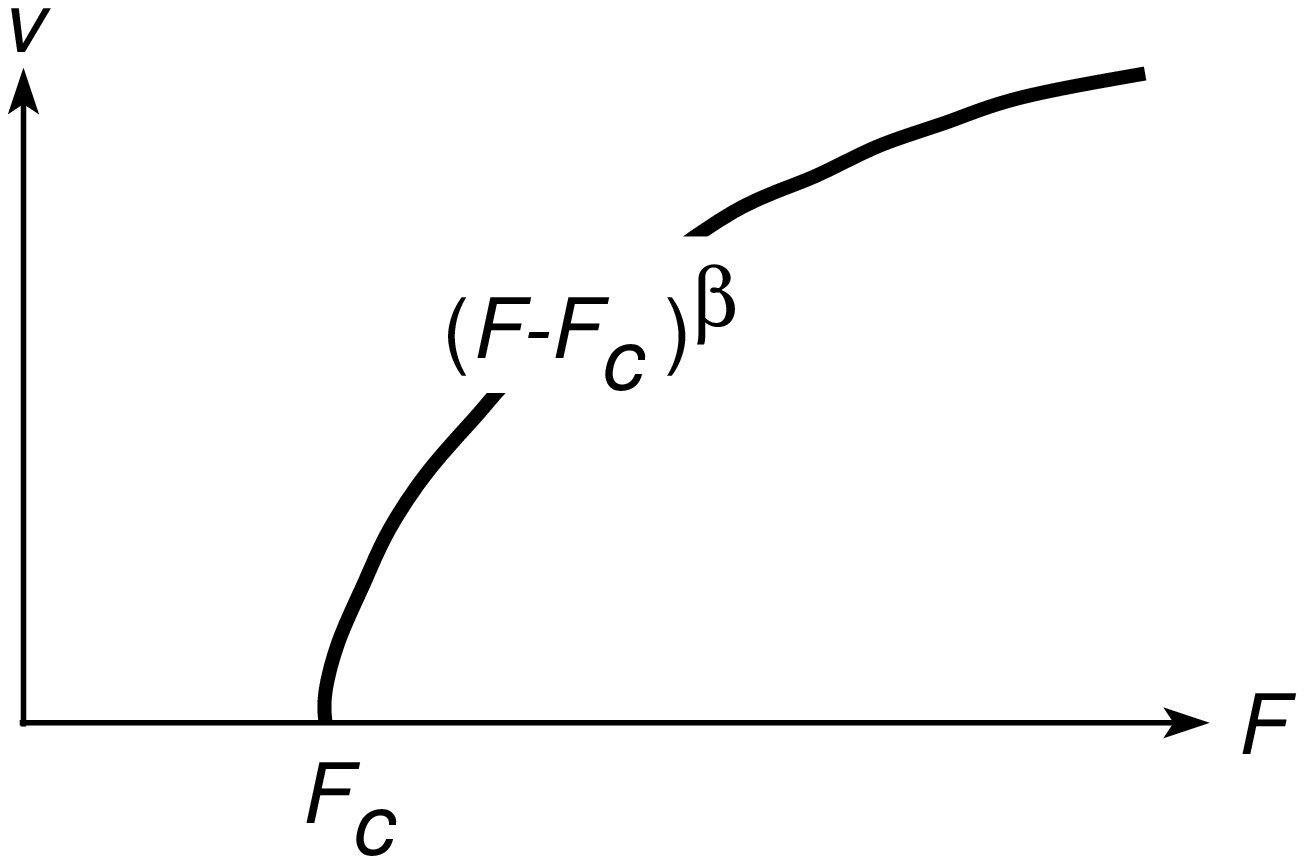}{Critical behavior of the velocity.}
When $F$ is small, the line is trapped in one of many metastable states in
which
$\partial r/\partial t=0$ at all points. For $F$ larger than a threshold $F_c$,
the line is depinned from the last metastable state, and moves with an average
velocity $v$. On approaching the threshold from above, the velocity vanishes as
\eqn\eVel{v=A(F-F_c)^\beta,}
where $\beta$ is the {\it velocity exponent}, and $A$ is a
nonuniversal amplitude. The motion just above threshold is not uniform,
composed of rapid jumps as large segments of the line depin from strong
pinning centers, superposed on the slower steady advance. These
jumps have a power law distribution in size, cutoff at a correlation length
$\xi$ which diverges at the transition as
\eqn\eXi{\xi\sim (F-F_c)^{-\nu}.}
The jumps are reminiscent of {\it avalanches} in other slowly driven systems.
In fact, the depinning can be approached from below $F_c$ by monotonically
increasing $F$ in small increments, each sufficient to cause a jump to the
next metastable state. The size and width of avalanches becomes invariant
on approaching $F_c$. For example,
\eqn\eAval{{\rm Prob(width\ of\ avalanche} > \ell) \approx{1\over \ell^\kappa}
\hat\rho(\ell/\xi_-),}
where the cutoff $\xi_-$ diverges as in Eq.\eXi.
The critical line is a {\it self--affine} fractal whose correlations satisfy
the dynamic scaling from
\eqn\eHcorr{\langle\left[r(x,t)-r(x',t')\right]^2\rangle
=(x-x')^{2\zeta}g\left({{|t-t'|}\over{|x-x'|^z}}\right),}
defining  the {\it roughness} and {\it dynamic} exponents, $\zeta$ and $z$
respectively. (Angular brackets reflect averaging over all realizations of
the random force $f$.) The scaling function $g$
goes to a constant as its argument approaches 0; $\zeta$ is
the wandering exponent of an instantaneous line profile, and
$z$ relates the average lifetime of an avalanche to its size
by $\tau(\xi)\sim\xi^z$.

Although, the underlying issues of collective depinning for CDWs
and interfaces have been around for some time, only recently a systematic
perturbative approach to the problem was developed. This functional
renormalization group (RG) approach to the dynamical equations of
motion was originally developed in the context of CDWs by Narayan and
Fisher\ref\rCDW{D.S.~Fisher, Phys. Rev. B {\bf 31}, 1396 (1985),
O.~Narayan and D.S.~Fisher, Phys. Rev. B {\bf 46}, 11520 (1992).}\  (NF),
and extended to interfaces by Nattermann et al\ref
\rNSTL{T.~Nattermann, S.~Stepanow, L.-H.~Tang, and
H.~Leschhorn, J. Phys. II France {\bf 2}, 1483 (1992).}.
We shall provide
a brief outline of this approach starting from Eq.\eGen. Before embarking
on the details of the formalism, it is useful to point out some scaling
relations
amongst the exponents which follow from underlying symmetries and
non-renormalization conditions.

\noindent{\bf 1.} As mentioned earlier, the motion of the line close to the
threshold is composed of jumps of segments of size $\xi$. Such jumps
move the interface forward by $\xi^\zeta$ over a time period $\xi^z$.
Thus the velocity behaves as,
\eqn\eVid{v\sim {\xi^\zeta \over \xi^z}\sim|F-F_c|^{\nu(z-\zeta)}\quad
\Longrightarrow \quad \beta=\nu(z-\zeta).}

\noindent{\bf 2.} If the elastic couplings are linear, the response of the line
to a
{\it static} perturbation $\varepsilon (x)$ is obtained simply by considering
\eqn\ernew{r_\varepsilon (x,t)=r(x,t)-{\cal K}^{-1}[\varepsilon (x)],}
where ${\cal K}^{-1}$ is the inverse kernel. Since, $r_\varepsilon $ satisfies
Eq.\eGen\ subject to a force $F+\varepsilon (x)+f(x,r_\varepsilon )$, $r$
satisfies
the same
equation with a force $F+f(x,r-{\cal K}^{-1}[\varepsilon (x)])$. As long as the
statistical properties of the stochastic force are not modified by the above
change in its argument, ${\partial \left\langle r \right\rangle}/\partial
\varepsilon =0$, and
\eqn\eSres{\left\langle {\partial {r_\varepsilon (x)}\over \partial \varepsilon
(x)}
\right\rangle=-{\cal K}^{-1},\quad {\rm or}\quad
\left\langle {\partial {r_\varepsilon (q)}\over \partial \varepsilon (q)}
\right\rangle
={1 \over |q|^\sigma}.}
Since it controls the macroscopic response of the line, the kernel ${\cal K}$
cannot change under RG scaling. From Eqs.\eHcorr\ and \eXi, we can
read off the scaling of $r(x)$, and the force $\delta F$, which using the
above non-renormalization must be related by the exponent relation
\eqn\eFid{\zeta+{1\over\nu}=\sigma.}
Note that this identity depends on the statistical invariance of noise under
the
transformation in Eq.\ernew. It is satisfied as long as the correlations
$\left\langle f(x,r)f(x',r') \right\rangle$ only depend on $r-r'$. The identity
does not hold if these correlations also depend on the slope
$ {\partial r/ \partial x}$.

\noindent {\bf 3.} A scaling argument related to the Imry--Ma estimate of the
lower critical dimension of the random field Ising model, can be used to
estimate the roughness exponent\ref
\rIM{Y. Imry and S.-K. Ma, Phys. Rev. Lett. {\bf 35}, 1399 (1975).}.
The elastic force on a segment of
length $\xi$ scales as $\xi^{\zeta-\sigma}$. If fluctuations in force are
uncorrelated in space, they scale as $\xi^{-(\zeta+1)/2}$ over the area of an
avalanche. Assuming that these two forces must be of the same order
to initiate the avalanche leads to
\eqn\eIMl{\zeta={2\sigma-1\over 3}.}
This last argument is not as rigorous as the previous two. Nonetheless,
all three exponent identities can be established within the RG framework.
Thus the only undetermined exponent is the dynamic one, $z$.

A field theoretical description of the dynamics of Eq.\eGen\ can be developed
using the formalism of Martin, Siggia and Rose\ref\rMSR{P.C.~Martin,
E.~Siggia, and H.~Rose, Phys. Rev. A {\bf 8}, 423 (1973).}\  (MSR):
Generalizing
to a $d-$dimensional interface, an auxiliary field
$\rh(\bx,t)$ is introduced to implement the equation of motion as a series of
$\delta$--functions. Various dynamical response and correlation functions for
the field $r(\bx,t)$ can then be generated from the functional,
\eqn\eZl{Z=\int{{\cal D} r(\bx,t){\cal D}\rh(\bx,t){\cal J}[r]\exp(S)},}
where
\eqn\eSl{S=i\int{d^d\bx\,\dt\rh(\bx,t)\left\{\partial_t r-{\cal K}[r]
-F-f\left({\bf x},r({\bf x},t)\right)\right\}}.}
The Jacobian ${\cal J}[r]$ is introduced to ensure that  the
$\delta$--functions
integrate to unity. It does not generate any new relevant terms and will be
ignored henceforth.

The disorder-averaged generating functional $\overline Z$ can be evaluated
by a saddle-point expansion
around a Mean-Field (MF) solution obtained by setting
$\K_{MF}[r(\bx)]=vt-r(\bx).$
This amounts to replacing interaction forces with Hookean springs connected
to the center of mass, which moves with a velocity $v$.
The corresponding equation of motion is
\eqn\eMF{{dr_{MF}\over dt}=vt-r_{MF}(t)+f[r_{MF}(t)]+F_{MF}(v),}
where the relationship $F_{MF}(v)$ between the external force $F$ and average
velocity $v$ is determined from the consistency condition
$\langle r_{MF}(t)\rangle=vt$.
The MF solution depends on the type of irregularity\rCDW:
For smoothly varying random potentials,
$\beta_{MF}=3/2$, whereas for cusped random potentials, $\beta_{MF}=1$.
Following the treatment of NF\rCDW$^,$\ref\rNF{O.~Narayan and D.~S.~Fisher,
Phys. Rev. B {\bf 48}, 7030 (1993).},
we use the mean field solution for cusped potentials,
anticipating jumps with velocity of $O(1)$, in which case
$\beta_{MF}=1$. After rescaling and averaging over impurity
configurations, we arrive at a generating functional whose
low-frequency form is
\eqn\eZav{\eqalign{
\overline Z &= \int{\cal D}R(\bx,t){\cal D}\Rh(\bx,t)\exp(\tilde S), \cr
\tilde S &= -\int \dx\dt\left[F-F_{MF}(v)\right] \hat R(\bx,t) \cr
  &\qquad-\int\dq\dw \hat R(-\bq,-\w)(-i\w\rho+|\bq|^\sigma)R(\bq,\w) \cr
  &\qquad+ {1 \over 2}\int\dx\dt dt'\, \hat R(\bx,t)\Rh(\bx,t')
 C\left[vt-vt'+R(\bx,t)-R(\bx,t')\right].\cr}}
In the above expressions, $R$ and $\Rh$ are coarse-grained
forms of $r-vt$ and $i\rh$, respectively. $F$ is adjusted to satisfy
the condition $\langle R\rangle=0$. The function $C(v\tau)$ is
initially the connected mean-field correlation function
$\langle(r_{MF}(t)r_{MF}(t+\tau)\rangle_c$.

Ignoring the $R$-dependent terms in the argument of $C$, the
action becomes Gaussian, and is invariant under a scale
transformation $x\to bx$, $t\to b^\sigma t$, $R\to b^{\sigma-d/2}R$,
$\hat R\to b^{-\sigma-d/2} \hat R$, $F\to b^{-d/2}F$, and
$v\to b^{-d/2}v$. Other terms in the action, of higher
order in $R$ and $\hat R$, that result from the expansion of $C$
[and other terms not explicitly shown in Eq.\eZav],
decay away at large length and time scales
if $d>d_c=2\sigma$. For $d>d_c$, the interface is smooth ($\zeta_0<0$)
at long length scales, and the depinning exponents take the Gaussian
values $z_0=\sigma$, $\nu_0=2/d$, $\beta_0=1$.

At $d=d_c$, the action $S$ has an infinite number of marginal
terms that can be rearranged as a Taylor series of the
marginal function $C\left[vt-vt'+R({\bf x},t)-R({\bf
x},t')\right]$, when $v\to 0$. The RG is carried out by
integrating over
a momentum shell $\Lambda/b<|\bq|<\Lambda$ (we set the
cutoff wave vector to $\Lambda=1$
for simplicity) and all frequencies, followed by a scale
transformation $x\to bx$, $t\to b^zt$, $R\to b^\zeta R$, and
$\hat R\to b^{\theta-d} \hat R$, where $b=e^\ell$.
The resulting recursion relation for the linear part
in the effective action (to all orders in perturbation theory) is
\eqn\eFrr{{\partial(F-F_{MF})\over \partial\ell}=(z+\theta)(F-F_{MF})+{\rm
constant},}
which immediately implies (with a suitable definition of $F_c$)
\eqn\eFcrr{{\partial(F-F_c)\over \partial\ell}= y_F(F-F_c),}
with the exponent identity
\eqn\eTid{y_F=z+\theta=1/\nu\quad.}
The functional renormalization of $C(u)$ in $d=2\sigma-\epsilon$
interface dimensions, computed to one-loop order,
gives the recursion relation,
\eqn\eCrr{\eqalign{
{\partial C(u) \over \partial \ell}=[\epsilon &+ 2\theta
+2(z-\sigma)]C(u)+\zeta u{dC(u)\over du} \cr
 &- {S_d\over (2\pi)^d}{d\over du}\left\{\left[C(u)-C(0)\right]
{dC(u)\over du}\right\},\cr}}
where $S_d$ is the surface area of a unit sphere in $d$ dimensions.
NF showed that all higher order diagrams contribute to the
renormalization of $C$ as
total derivatives with respect to $u$, thus, integrating
Eq.\eCrr\
at the fixed-point solution $\partial C^*/\partial\ell = 0$,
together with Eqs.\eFid\ and \eTid,
gives $\zeta=\epsilon/3$ to all orders in $\epsilon$,
provided that $\int C^*\neq 0$. This gives Eq.\eIMl\ for a
one-dimensional interface, as argued earlier.
This is a consequence of the fact that $C(u)$ remains short-ranged
upon renormalization, implying the absence of anomalous
contributions to $\zeta$.

The dynamical exponent $z$ is calculated through the renormalization
of $\rho$, the term proportional to $\Rh\partial_tR$, which yields
\eqn\eZexp{z=\sigma-2\epsilon/9+O(\epsilon^2),}
and using the exponent identity \eVid,
\eqn\eBexp{\beta=1-2\epsilon/9\sigma+O(\epsilon^2).}
Nattermann et. al.\rNSTL\ obtain the same
results to $O(\epsilon)$ by directly averaging the MSR generating
function in Eq.\eZl, and expanding perturbatively around a rigidly
moving interface.

Numerical integration of Eq.\eGen\ for an elastic interface\ref\rDong{
M.~Dong, M.~C.~Marchetti, A.~A.~Middleton, and V.~Vinokur, Phys. Rev.
Lett. {\bf 70}, 662 (1993). The identification of the exponent $\zeta=1$ from
correlation function has been questioned by
H. Leschhorn and L.-H. Tang, Phys. Rev. Lett. {\bf 70}, 2973 (1993). }\
$(\sigma=2)$
has yielded critical exponents $\zeta=0.97\pm0.05$ and $\nu=1.05\pm0.1$,
in agreement with the theoretical result $\zeta=\nu=1$. The velocity
exponent $\beta=0.24\pm0.1$ is also consistent with the one-loop theoretical
result 1/3; however, a logarithmic dependence $v\sim1/\ln(F-F_c)$, which
corresponds to $\beta=0$, also describes the numerical data well.
In contrast, experiments and various discrete models of interface growth
have resulted in scaling behaviors that differ from system to system.
A number of different experiments
on fluid invasion in porous media\ref\rIntexp{M.A. Rubio, C.A. Edwards,
A. Dougherty, and J.P Gollub, Phys. Rev. Lett. {\bf 63}, 1685 (1989);
V.K. Horv\'ath, F. Family, and T. Vicsek, Phys. Rev. Lett. {\bf 67},
3207 (1991); S.~He, G.~L.~M.~K.~S.~Kahanda, and P.-Z. Wong, Phys. Rev.
Lett. {\bf 69}, 3731 (1992).}\
give roughness exponents of around 0.8,
while imbibition experiments\ref\rBul{S.~V.~Buldyrev,
A.-L.~Barabasi, F.~Caserta, S.~Havlin, H.~E.~Stanley, and T.~Vicsek,
Phys. Rev. A {\bf 45}, R8313 (1992).}$^,$\ref
\eFCA{F.~Family, K.~C.~B.~Chan,
and J.~G.~Amar, in {\it Surface Disordering: Growth, Roughening and
Phase Transitions}, Les Houches Series, Nova Science Publishers,
New York (1992).}\
have resulted in $\zeta\approx0.6$.
A discrete model studied by Leschhorn\ref\rHeiko{H.~Leschhorn,
Physica A {\bf 195}, 324 (1993).}, motivated by Eq.\eGen\ with $\sigma=2$,
gives a roughness exponent of 1.25 at threshold.
Since the expansion leading to Eq.\eGen\
breaks down when $\zeta$ approaches one, it is not clear how to reconcile
the results of Leschhhorn's numerical work\rHeiko\ with the coarse-grained
description of the RG calculation, especially since any model with $\zeta>1$
cannot have a coarse grained description based on gradient expansions.

Amaral, Barabasi, and Stanley (ABS)\ref\rAmar{L.~A.~N.~Amaral,
A.-L.~Barabasi, and  H.~E.~Stanley, Phys. Rev. Lett. {\bf 73}, 62 (1994).}\
recently pointed out
that various models of interface depinning in 1+1 dimensions fall
into two distinct classes, depending on the tilt dependence of
the interface velocity:

\noindent {\bf 1.} For models like the random field Ising Model\ref
\rRFIM{H.~Ji and M.~O.~Robbins, Phys. Rev. B {\bf 44}, 2538 (1991);
B.~Koiller, H.~Ji, and M.~O.~Robbins, Phys. Rev. B {\bf 46}, 5258 (1992).},
and some Solid On Solid models, the computed exponents are consistent
with the exponents given by the RG analysis. It has been suggested\rHeiko,
however, that the roughness exponent is systematically larger than
$\epsilon/3$, casting doubt on the exactness of the RG result.

\noindent {\bf 2.} A number of different models, based on
directed percolation (DP)\ref\rTL{L.-H.~Tang and H.~Leschhorn,
Phys. Rev. A {\bf 45}, R8309 (1992).}$^,$\rBul\ give a
different roughness exponent, $\zeta\approx0.63$. In these
models, pinning sites are randomly distributed with a probability
$p$, which is linearly related to the force $F$. The interface
is stopped by the boundary of a DP cluster of pinning sites. The
critical exponents at depinning can then be related to the
longitudinal and transverse correlation length exponents
$\nu_\parallel\approx1.70$ and $\nu_\perp\approx1.07$ of DP.
In particular, $\zeta=\nu_\parallel/\nu_\perp\approx0.63$, and
$\beta=\nu_\parallel-\nu_\perp\approx0.63$, in
agreement with experiments.

The main difference of these models can be understood in terms
of the dependence of the threshold force $F_c$ to the orientation.
To include the possible dependence of the line mobility on its slope,
$\partial_x r$, we can generalize the equation of motion to
\eqn\etiltline{\partial_t r=K\partial_x^2 r + \kappa \partial_x r +
{\lambda\over2} (\partial_x r)^2 + F + f(x,r).}
The isotropic depinning studied by RG corresponds to $\kappa=\lambda=0$.
In models of depinning by directed percolation studied so far\rTL\rsc\rBul\
there is a dependence of $F_c$ on slope, making a nonzero $\lambda$
unavoidable.  The nonlinearity is relevant, accounting for the different
universality
class. Eq.\etiltline\ with $\kappa=0$, motivated in a different fashion, has
been
studied by Stepanow\ref
\rStepanow{S.~Stepanow, preprint (1993).}.
The exponents obtained approximately by a one loop expansion,
$\zeta\approx 0.8615$, $z=1$, and $y_F\approx 0.852$ are reasonably
close to those of directed percolation.
The presence of anisotropy in depinning actually suggests
a third possibility:

\noindent {\bf 3.} When the line is depinning along a (tilted)
direction of lower symmetry, even more relevant terms like
$\kappa\partial_x r$ will be present in the equation of motion.
This new universality class is possibly controlled by
``tilted" DP clusters\ref\rTang{L.-H. Tang, M. Kardar, and D. Dhar,
work in progress.}, for which $\zeta=1/2$.

For the case of the contact line (CL) $(\sigma=1)$, these anisotropies are
irrelevant, but there are other concerns related to the details of the
driving force: In most experiments, the velocity of the CL is controlled rather
than the external force. The effect of this can be numerically investigated
by replacing the external force $F$ in Eq.\eGen\ with
\eqn\eCv{F'=v-\int {dx'\over L} f(x',r(x',t)),}
and looking at the time average of $F'$ as a function of $v$.
($F'$ is chosen such that $\int dx \del_t r(x)=vL$.)
Even though the critical behavior for both ways of driving may be
the same for an infinitely large system, there is a system size dependent
region near the depinning threshold where the behavior changes drastically.
Preliminary findings on an elastic line suggest that in this region,
the velocity exponent $\beta$ becomes considerably larger than one, in
marked contrast with the constant force case. This can be qualitatively
understood as follows: For a system of finite size, when a constant
driving force is applied, the average velocity
drops to zero as soon as temporal fluctuations of the instantaneous
velocity are comparable with the time-averaged velocity. This is
because the time average is then completely dominated by configurations
for which the interface is pinned. Thus, the
pinning transition becomes truly second order only in the large
system limit: The velocity jumps to zero from a finite value in
a finite system. In contrast, for constant velocity driving, no
configuration has more weight than any other, since the interface is
constrained to move past any obstacles by suitably increasing the
applied external force, and decreasing it when passing through weakly
pinning regions. Thus, in the region where a force-driven
interface is pinned, the velocity-driven interface will experience
fluctuations in the external force comparable to the average force
itself. This average force as a function of velocity has an
effective velocity exponent much larger than one.
This distinction may partially explain
the large velocity exponent found in a recent
CL experiment\ref\rStokes{J.P.~Stokes,
M.~J.~Higgins, A.~P.~Kushnick, S.~Bhattacharya, and M.~O.~Robbins,
Phys. Rev. Lett. {\bf 65}, 1885 (1990).}, where the
interface was velocity-driven. In addition to this,
gravity imposes a finite wavelength cutoff on the roughening of the CL,
which may complicate the analysis of experimental results.

\medskip
%% LECTURE 2
\vskip .15truein \noindent{\bf II. DEPINNING OF A LINE IN THREE
DIMENSIONS}\smallskip\par

The pinning of flux lines (FLs) in Type-II superconductors is of
fundamental importance to many technological applications that
require large critical currents\ref\rreview{See, for example,
G.~Blatter et. al., ETH preprint, and references therein.}.
Upon application of an external current density ${\bf J}$,
the FL becomes subject to a Lorentz force per unit length
\eqn\eLF{\bF=\frac{J\phi_0}{c}\,{\bf \hat J}\times{\bf \hat t},}
where $\phi_0$ is the flux quantum, and ${\bf \hat t}$ is the unit
tangent vector along the FL, which points along the local magnetic
field.
The motion of FLs due to the Lorentz force causes undesirable
dissipation of supercurrents. Major increases in the critical current
density $J_c$ of a sample are achieved when the FLs are pinned
to impurities.

Recent numerical simulations have concentrated on the low temperature
behavior of a single FL near depinning\ref\rEnomoto{Y.~Enomoto,
Phys. Lett. A {\bf 161}, 185 (1991); Y.~Enomoto, K.~Katsumi, R.~Kato,
and S.~Maekawa, Physica C {\bf 192}, 166 (1992).}\rsc\rDong\rsc\ref
\rFeng{C.~Tang, S.~Feng, and L.~Golubovic, Phys. Rev. Lett. {\bf 72}, 1264
(1994).},
mostly ignoring fluctuations transverse
to the plane defined by the magnetic field and the Lorentz force.
Common signatures of the depinning transition from $J<J_c$ to
$J>J_c$ include a broad band ($f^{-a}$ type) voltage noise spectrum,
and self-similar fluctuations of the FL profile.

\epsfysize=7cm\figure\fFL{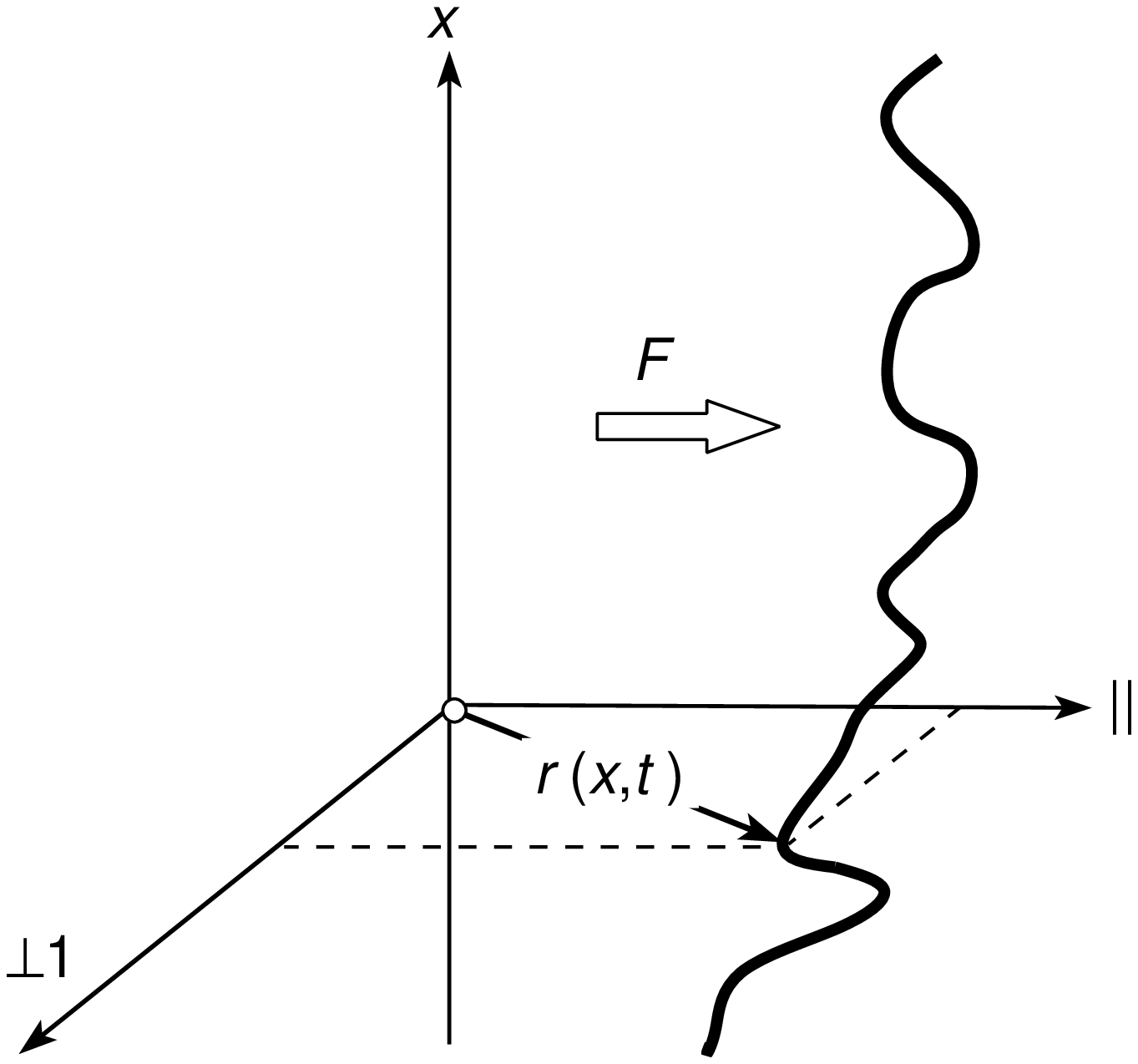}{Geometry of the line in three dimensions.}
The configuration of the FL at time $t$ is now described by the
vector function $\br(x,t)$,
where $x$ is along the magnetic field {\bf B}, and the unit vector
${\bf e}_\parallel$ is along the Lorentz force $\bF$.
(See Fig.\fFL)
The major difference of the FL from the line in two dimensions
is that the position, $\br(x,t)$, is now a
2-dimensional vector instead of a scalar; fluctuating along both
${\bf e}_\parallel$ and ${\bf e}_\perp$ directions.
Point impurities are modeled by a random
potential $V(x,\br)$, with zero mean and short-range correlations.
The simplest possible Langevin equation for the FL, consistent with
{\it local, dissipative dynamics}, is
\eqn\eMot{
\rho\frac{\partial \br}{\partial t} =
\partial_x^2 \br+\bfr \left(x,\br(x,t)\right)+\bF,
}
where $\rho$ is the inverse mobility of the FL, and $\bfr=-\nabla_\br V$.
The potential $V(x,\br)$ need not be
isotropic. For example, in a single crystal of ceramic
superconductors
with the field along the oxide planes, it will be easier to move
the FL along the planes. This leads to a pinning threshold that
depends on the orientation of the force. Anisotropy also modifies the
line tension, and the elastic term in Eq.\eMot\ is in general
multiplied by a non-diagonal matrix $K_{\alpha\beta}$.
The random force $\bfr(x,\br)$, can be taken to have zero
mean with correlations
\eqn\estat{
\langle f_\alpha(x,\br)f_\gamma(x',\br')\rangle=
\delta(x-x')\Delta_{\alpha\gamma}(\br-\br').
}
We shall focus mostly on the isotropic case,
with $\Delta_{\alpha\gamma}(\br-\br')=
\delta_{\alpha\gamma}\Delta(|\br-\br'|)$,
where $\Delta$ is a function that decays rapidly for large
values of its argument.

In addition to the exponents defined in the first lecture,
$(\beta,\ \nu,\ \zeta\to\zetal,\ z\to\zl)$, there are two additional
critical exponents that describe
fluctuations transverse to the overall motion of the FL
slightly above depinning. At length scales
up to $\xi$, the correlated fluctuations satisfy the dynamic
scaling form,
\eqn\escaling{\left\{\eqalign{
\langle[r_\parallel(x,t)-r_\parallel(x',t')]^2\rangle
&= |x-x'|^{2\zeta_\parallel}g_\parallel\left(
\frac{|t-t'|}{|x-x'|^{z_\parallel}}\right),\cr
\langle[r_\perp(x,t)-r_\perp(x',t')]^2\rangle
&= |x-x'|^{2\zeta_\perp}g_\perp\left(
\frac{|t-t'|}{|x-x'|^{z_\perp}}\right),\cr }\right.
}
where $\zeta_\perp$ and $z_\perp$ are the transverse roughness and
dynamic exponents.
One consequence of transverse fluctuations is that a ``no passing"
rule\ref\rMiddleton{A.~A.~Middleton and D.~S.~Fisher, Phys. Rev. Lett.
{\bf 66}, 92 (1991); Phys. Rev. B {\bf 47}, 3530 (1993).}, applicable to
CDWs and interfaces, does not apply to FLs. It is possible to have
coexistence of
moving and stationary FLs in particular realizations of the random
potential.
The effects of transverse fluctuations $r_\perp$ for large driving
forces, when
the impurities act as white noise, will be discussed later.
At this point, we would like to know how these
transverse fluctuations scale near the depinning transition, and whether
or not they influence the critical dynamics of longitudinal fluctuations
near threshold.

The answer to the second question is obtained by the following
qualitative argument:
Consider Eq.\eMot\ for a particular realization of
randomness
$\bfr(x,\br)$. Assuming that portions of the FL always
move in the forward direction\ref\rback{This is not strictly true,
but backward motion happens
very rarely, and only at short length scales.}, there is a unique
point $r_\perp(x,r_\parallel)$ that is visited by the line
for given coordinates $(x,r_\parallel)$. We construct a new force
field $f'$ on a two dimensional space $(x,r_\parallel)$
through $f'(x,r_\parallel)\equiv f_\parallel\left(x,r_\parallel,
r_\perp(x,r_\parallel)\right)$. It is then clear that the dynamics
of the longitudinal component $r_\parallel(x,t)$ in a given force
field
$\bfr(x,\br)$ is identical to the dynamics
of $r_\parallel(x,t)$ in a force field $f'(x,r_\parallel)$, with
$r_\perp$ set to zero. It is quite plausible that, after
averaging over all $\bfr$, the correlations in $f'$ will
also be short-ranged, albeit different from those of
$\bfr$. Thus, the scaling of longitudinal fluctuations of the
depinning FL will not change upon taking into account transverse
fluctuations. However, the question of how these transverse
fluctuations scale still remains.

Certain statistical symmetries of the system restrict the form of
response and correlation functions. For example, Eq.\eMot\
has statistical space- and time-translational invariance, which
enables
us to work in Fourier space, i.e. $(x,t)\to(q,\omega)$.
For an {\it isotropic} medium, $\bF$ and $\bv$ are parallel to
each other, i.e., $\bv(\bF)=v(F){\bf \hat F}$, where ${\bf \hat F}$
is
the unit vector along \bF. Furthermore, all
expectation values involving odd powers of a transverse
component are identically zero due to the statistical invariance
under the
transformation $r_\perp\to -r_\perp$. Thus, linear
response and two-point correlation functions  are {\it diagonal}.
The introduced critical exponents are then related through scaling
identities.
These can be derived from the
linear response to an infinitesimal external force field
$\beps(q,\omega)$,
\eqn\echi{
\chi_{\alpha\beta}(q,\omega)=\left<\frac{\partial r_\alpha(q,\omega)}
{\partial\varepsilon_\beta(q,\omega)}\right>\equiv
\delta_{\alpha\beta}
\chi_\alpha,
}
in the $(q,\omega)\to(0,0)$ limit. Eq.\eMot\
is statistically invariant under the transformation $\bF\to\bF+
\beps(q)$,
$\br(q,\omega)\to\br(q,\omega)+q^{-2}\beps(q)$.
Thus, the static linear response has the form
$\chi_\parallel(q,\omega=0)=\chi_\perp(q,\omega=0)=q^{-2}$.
Since $\varepsilon_\parallel$ scales like the applied force,
the form of the linear response at the correlation length $\xi$
gives an exponent identity similar to Eq.\eFid :
\eqn\enuexp{
\zeta_\parallel+1/\nu=2.
}
Considering the transverse linear response
seems to imply $\zeta_\perp=\zeta_\parallel$. However,
the static part of the transverse linear response is
irrelevant at the critical RG fixed point, since
$z_\perp>z_\parallel$, as shown below.
When a slowly varying uniform external force $\beps(t)$ is applied,
the FL responds as if the instantaneous external force
$\bF+\beps$ is a constant, acquiring an average velocity,
\eqn\eDyn{
\langle\partial_t r_\alpha\rangle=v_\alpha(\bF+\beps)\approx
v_\alpha(\bF)+\frac{\partial v_\alpha}{\partial F_\gamma}
\varepsilon_\gamma.
}
Substituting $\partial v_\parallel/\partial F_\parallel=dv/dF$ and
$\partial v_\perp/\partial F_\perp=v/F$, and Fourier transforming,
gives
\eqn\eChi{\eqalign{
\chi_\parallel(q=0,\omega) &= \frac{1}{-i\omega(dv/dF)^{-1}
+O(\omega^2)}, \cr
\chi_\perp(q=0,\omega) &= \frac{1}{-i\omega(v/F)^{-1}+O(\omega^2)}.\cr}
}
Combining these with the static response, we see that the
characteristic
relaxation times of fluctuations with wavelength $\xi$ are
\eqn\eTim{\eqalign{
\tau_\parallel(q=\xi^{-1}) &\sim \left(q^2\frac{dv}{dF}\right)^{-1}
\sim\xi^{2+(\beta-1)/\nu}\sim\xi^{z_\parallel}, \cr
\tau_\perp(q=\xi^{-1}) &\sim \left(q^2\frac{v}{F}\right)^{-1}
\sim\xi^{2+\beta/\nu}\sim\xi^{z_\perp}, \cr}
}
which, using Eq.\enuexp\ , yield the scaling relations
\eqn\eBetaid{\eqalign{
\beta &=(z_\parallel-\zeta_\parallel)\nu, \cr
z_\perp &= z_\parallel+1/\nu. \cr }
}
We already see that the dynamic relaxation of transverse
fluctuations is much slower than longitudinal ones.
All critical exponents can be calculated from $\zeta_\parallel$,
$\zeta_\perp$, and $z_\parallel$, by using Eqs.\enuexp\ and
\eBetaid.

Equation \eMot\ can again be analyzed using the MSR formalism.
The long wavelength, low frequency behavior, for isotropic
random potentials, is described by the effective action
\eqn\eSnew{\eqalign{
{\tilde S}= &-\int\dt\dx [F-F_{MF}(v)]\Rh_\parallel(\bx,t) \cr
&-\int\dq\dw\Rh_\parallel(-\bq,-\omega)
R_\parallel(\bq,\omega)(-i\omega\rho+q^2) \cr
&-\int\dq\dw\bRh_\perp(-\bq,-\omega)\cdot\bR_\perp(\bq,\omega)
\left(-i\omega\frac{F_c}{v}+q^2\right) \cr
&+\frac{1}{2}\sum_\gamma\int\dx\dt dt'\,\Rh_\gamma(\bx,t)
\Rh_\gamma(\bx,t') C_\gamma\left(v(t-t')
+R_\parallel(\bx,t)-R_\parallel(\bx,t')\right).\cr
}}
All terms in $\tilde S$ involving longitudinal fluctuations
are identical to the
two-dimensional case, thus we obtain the same critical exponents for
longitudinal fluctuations, i.e., $\zeta_\parallel=\epsilon/3$,
$z_\parallel=2-2\epsilon/9+O(\epsilon^2)$.
The renormalization of transverse temporal force-force
correlations $C_\perp(u)$
yields an additional recursion relation
\eqn\eCperp{\eqalign{
\frac{\partial C_\perp(u)}{\partial\ell} &=
[\epsilon+2\theta_\perp+2(z_\parallel-2)]C_\perp(u)
+\zeta_\parallel u\,\frac{dC_\perp(u)}{du} \cr
\noalign{\medskip}
&\qquad-\frac{S_d}{(2\pi)^d}\left\{[C_\parallel(u)-C_\parallel(0)]
\frac{d^2C_\perp(u)}{du^2}\right\}.
}}
In addition, the form of the transverse dynamic linear response
given in Eq.\eChi\ implies the nonrenormalization of the term
proportional to $\Rh_\perp\del_t R_\perp(F_c/v)$, which, along with the
renormalization of $C_\perp(0)$, gives a transverse roughness exponent
$\zeta_\perp=\zeta_\parallel-d/2$, to all orders in perturbation
theory.
For the FL $(\epsilon=3)$, the critical exponents are then given by
\eqn\eFLexp{\eqalign{
\zeta_\parallel&=1,\; z_\parallel\approx4/3,\;  \nu=1, \cr
\beta&\approx 1/3,\;  \zeta_\perp=1/2,\;  z_\perp\approx7/3. \cr}
}

\epsfysize=5cm\figure\fVE{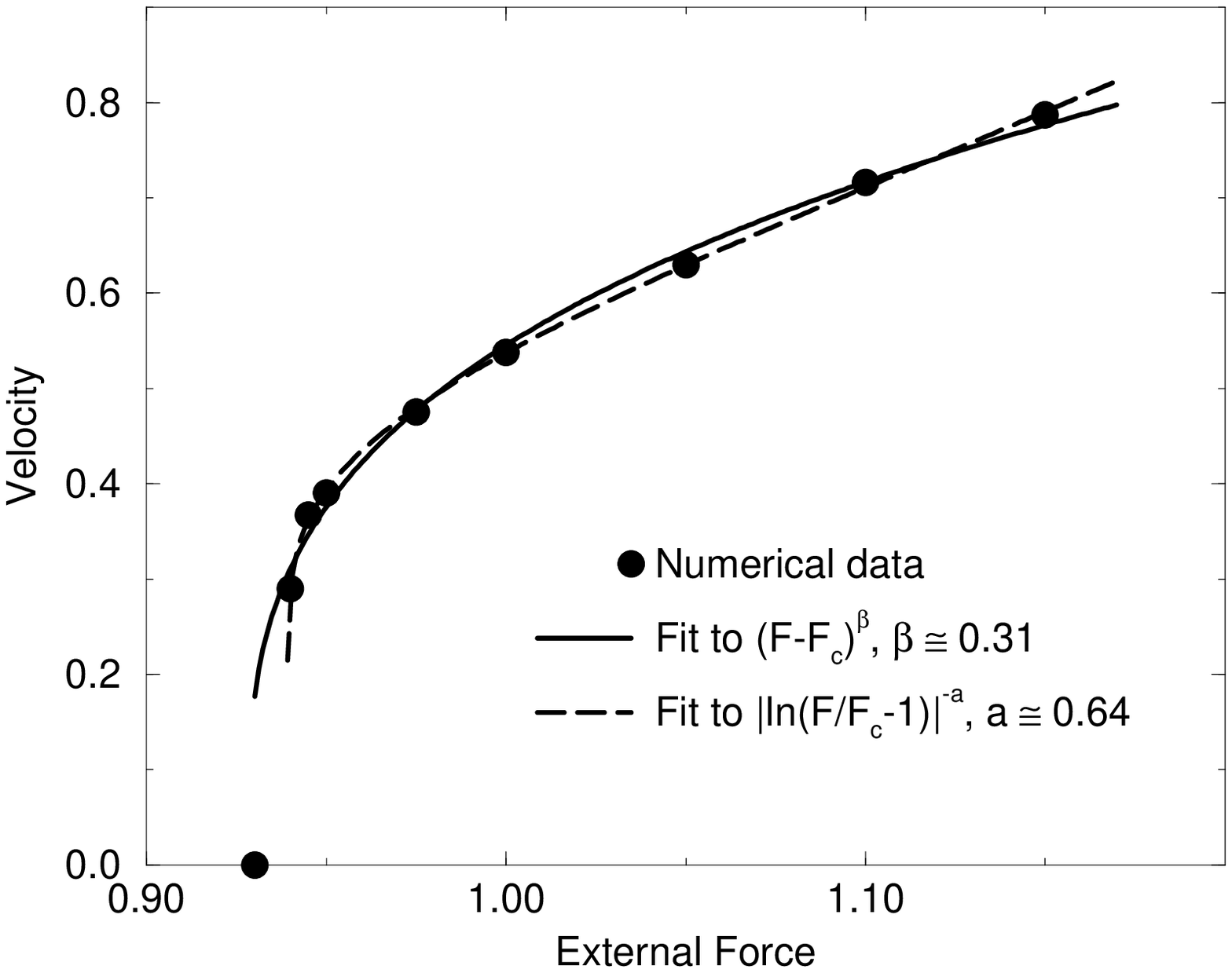}{A plot of average velocity versus
external force for a system of 2048 points. Statistical errors are smaller
than symbol sizes. Both fits have three adjustable parameters: The
threshold force, the exponent, and an overall multiplicative constant.}

\epsfysize=5cm\figure\fRE{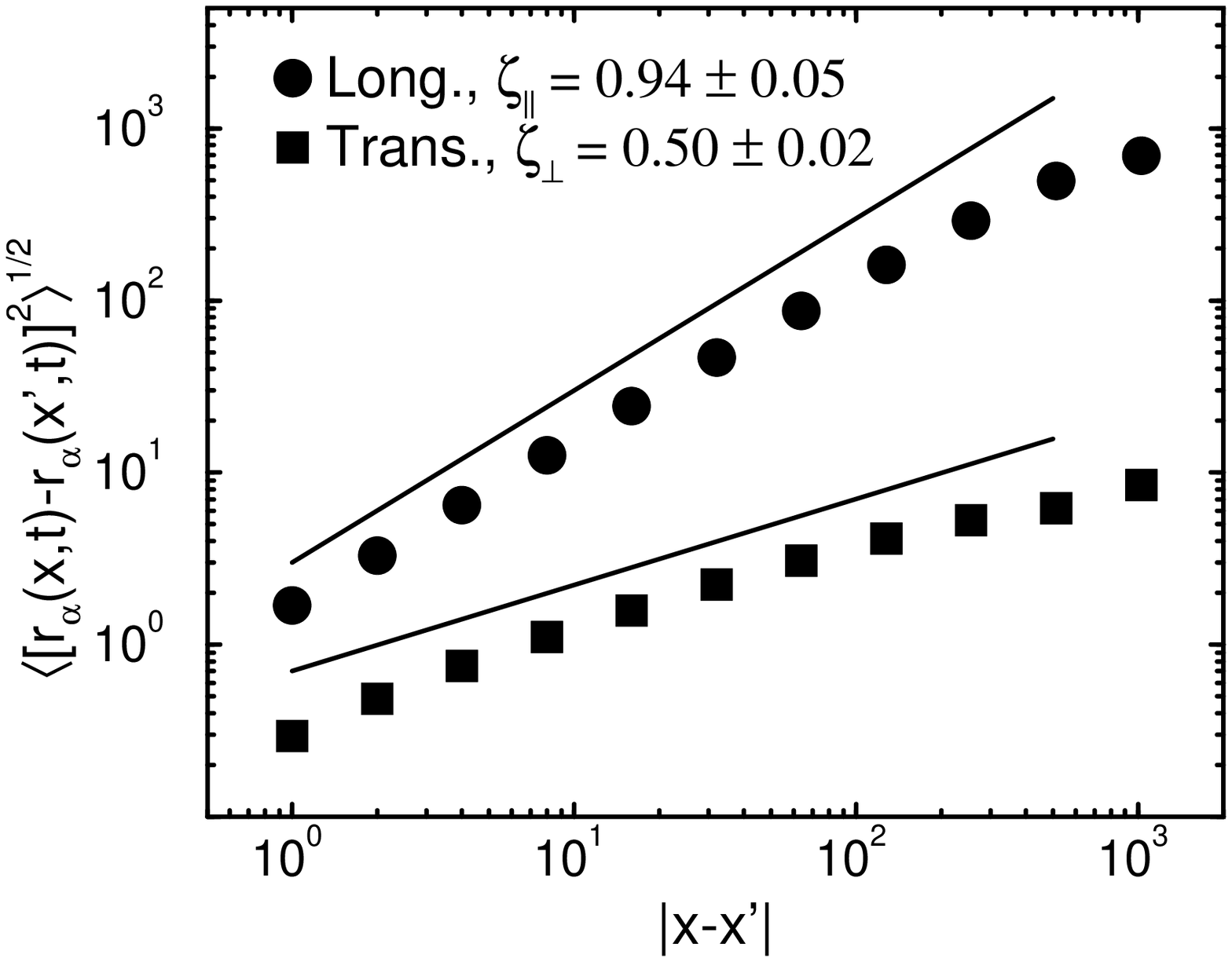}{A plot of equal time correlation
functions versus separation, for the system shown in Fig.\fVE,
at $F=0.95$.
The observed roughness exponents very closely follow the theoretical
predictions of $\zeta_\parallel=1,\;\zeta_\perp=0.5$, which are shown
as solid lines for comparison.}

Numerical integrations  of Eq.\eMot\ref\rEKfluxdepin{D.~Erta\c s and
M.~Kardar, preprint (1994).}\ that test the scaling forms and
exponents predicted by Eqs.\eVel\ and \escaling\ are in agreement
with RG results: A fit for the velocity exponent gives $\beta=0.3\pm0.1,$
although a logarithmic fit $(\beta=0$) cannot be ruled out, as seen in
Fig.\fVE. The roughness exponents (see Fig.\fRE) fit the scaling form well,
with\rDong\
$\zeta_\parallel=0.94\pm0.05$, and
$\zeta_\perp=0.50\pm0.02$.

The potential pinning the FL in a single superconducting crystal is
likely to be highly {\it anisotropic}. For example, consider a
magnetic field parallel to the copper oxide planes of a ceramic
superconductor. The threshold force then depends on its orientation,
with depinning easiest along the copper oxide planes.
In general, the average velocity may depend on the orientations
of the external force and the FL. The most general gradient
expansion for the equation of motion is then,
\eqn\eAnis{
\frac{\del r_\alpha}{\del t} = \mu_{\alpha\beta}F_\beta +
\kappa_{\alpha\beta}\del_x r_\beta +K_{\alpha\beta}\del_x^2 r_\beta
+\frac{1}{2}\Lambda_{\alpha,\beta\gamma}\del_xr_\beta\del_xr_\gamma
+f_\alpha\left(x,\br(x,t)\right)+\cdots,}
with
\eqn\eAnisnoise{
\langle f_\alpha(x,\br)f_\beta(x',\br')\rangle
= \delta(x-x')C_{\alpha\beta}(\br-\br').}
Depending on the presence or absence of various terms allowed by
the symmetries of the system, the above set of equations encompasses
many distinct universality classes. For example, consider the situation
where \bv\ depends on \bF,
but not on the orientation of the line.
Eqs.\echi\  and \eChi\ have to be modified, since \bv\ and \bF\ are no
longer parallel (except along the axes with
$\br\to -\br$ symmetry), and the linear response function is not
diagonal. The RG analysis is more cumbersome:
For depinning along a non-symmetric direction,
the longitudinal exponents are not modified (in agreement
with the argument presented earlier), while the transverse
fluctuations are further suppressed to $\zeta_\perp=
2\zeta_\parallel-2$ (equal to zero for
$\zeta_\parallel=1$)\ref\rfootnote{In this case,
the longitudinal direction is chosen to be
along the average velocity \bv, not the Lorentz force \bF.}.
Relaxation of transverse modes are still characterized by
$z_\perp=z_\parallel+1/\nu$, and the exponent identity \enuexp\
also holds. Surprisingly, the exponents for depinning along
axes of reflection symmetry are the same as the isotropic case.
If the velocity also depends on the tilt,
there will be
additional relevant terms in the MSR partition function, which
invalidate the arguments leading to Eqs.\enuexp--\eBetaid.
The analogy to FLs in a planes
suggests that the longitudinal exponents for $d=1$ are controlled by
DP clusters\rTL$^,$\rBul, with
$\zeta_\parallel\approx 0.63$. Since no
perturbative fixed point is present in this case, it is not
clear how to explore the behavior of transverse fluctuations
systematically.
%\eject
%%LECTURE 3
\vskip .15truein \noindent{\bf III. NONLINEAR DYNAMICS OF MOVING
LINES}\smallskip\par

We have so far investigated the dynamics of a line near the depinning
transition. Now, we would like to consider its behavior in a different
regime, when the external driving force is large, and the impurities
appear as weak barriers that deflect portions of the line
without impeding its overall drift. In such non--equilibrium systems,
one can regard the evolution equations as more fundamental,
and proceed by constructing the most general equations consistent with the
symmetries and conservation laws of the situation under study\ref
\rMK{M. Kardar, in {\it Disorder and Fracture}, edited by J.C. Charmet,
S. Roux, and E. Guyon, Plenum, New York (1990); T. Hwa and M. Kardar, Phys.
Rev. A {\bf 45}, 7002 (1992).}.
Even in a system with isotropic randomness, which we will discuss here,
the average drift velocity, $v$, breaks the symmetry between
forward and backward motions, and allows introduction of
nonlinearities in the equations of motion\ref
\rPRL{M. Plischke, Z. R\'acz, and D. Liu, Phys. Rev. B {\bf 35}, 3485 (1987).}
$^,$\rMK.

Let us first concentrate on an interface in two dimensions. (Fig.\CL.)
By contracting up to two spatial derivatives of $r$, and keeping terms
that are relevant, one obtains
the Kardar-Parisi-Zhang\ref
\rKPZ{M. Kardar, G. Parisi, and Y. Zhang, Phys. Rev. Lett.
{\bf 56}, 889 (1986).}\ (KPZ) equation,
\eqn\eKPZ{
\del_tr(x,t)=\mu F+K\del_x^2r(x,t)+\frac{\lambda}{2}
\left[\del_xr(x,t)\right]^2+f(x,t),}
with random force correlations
\eqn\enoise{\langle f(x,t)f(x',t')\rangle=
2T\delta(x-x')\delta(t-t').}
For a moving line, the term proportional to the external force can
be absorbed without loss of generality by considering a suitable
Galilean transformation, $r\to r-at$, to a moving frame.
A large number of stochastic nonequilibrium growth models,
like the Eden Model and various ballistic deposition models
are known to be well described, at large length scales and times,
by this equation, which is intimately related to several other
problems. For example,
the transformation $v(x,t)=-\lambda \del_x r(x,t)$ maps Eq.\eKPZ\
to the randomly stirred {\it Burgers' equation} for fluid flow\ref
\rJMB{J.M. Burgers, {\it The Nonlinear Diffusion Equation} (Riedel,
Boston, 1974).}\rsc\ref\rFNS{D.~Forster, D.~R.~Nelson, and
M.~J.~Stephen, Phys. Rev. A {\bf 16}, 732 (1977).},
\eqn\eBurg{\del_tv+v\del_xv=K\del_x^2v-\lambda\del_xf(x,t).}

The correlations of the line profile still satisfy
the dynamic scaling form in Eq.\eHcorr, nevertheless with different
scaling exponents $\zeta, z$ and scaling function $g$. This self-affine
scaling is not critical, i.e., not obtained by fine tuning an external
parameter like the force, and is quite different in nature than the
critical scaling of the line near the depinning transition, which
ceases beyond the correlation length scale $\xi$.

Two important nonperturbative properties of Eq.\eKPZ\ help us determine
these exponents exactly in 1+1 dimensions:

\noindent{\bf 1.} {\it Galilean Invariance (GI):} Eq.\eKPZ\ is statistically
invariant
under the infinitesimal reparametrization
\eqn\eGI{r'=r+\epsilon x~,~x'=x+\lambda\epsilon t~,~t'=t,}
provided that the random force $f$ does not have temporal correlations\ref
\rMHKZ{E.~Medina, T.~Hwa, M.~Kardar, and Y.~Zhang, Phys. Rev. A
{\bf 39}, 3053 (1989).}.
Since the parameter $\lambda$ appears both in the transformation and
Eq.\eKPZ, it is not renormalized under any RG procedure that preserves
this invariance.
This implies the exponent identity\rFNS$^,$\rMHKZ
\eqn\eGIid{\zeta+z=2.}

\noindent{\bf 2.} {\it Fluctuation--Dissipation (FD) Theorem:} Eqs.\eKPZ\ and
\enoise\
lead to a Fokker--Planck equation for the evolution of
the joint probability ${\cal P}\left[r(x)\right]$,
\eqn\eFP{
\partial_t{\cal P}=\int dx\,\left( {\delta {\cal P}\over
\delta r(x)}\,\partial_t r + T{\delta^2{\cal P}
\over [\delta r(x)]^2}\right).
}
It is easy to check that ${\cal P}$ has a stationary solution
\eqn\eSS{{\cal P}=\exp\left(-{K\over 2T}\int dx\,(\del_x r)^2\right).}
If ${\cal P}$ converges to this solution,
the long--time behavior of the correlation
functions in Eq.\eHcorr\ can be directly read off Eq.\eSS,
giving $\zeta=1/2$.

Combining these two results, the roughness and dynamic exponents are
exactly determined for the line in two dimensions as
\eqn\eKPZexp{\zeta=1/2~,\qquad z=3/2.}
Many direct numerical simulations and discrete growth models have
verified these exponents to a very good accuracy. Exact exponents are
not known for interfaces in higher dimensions, since the FD property
is only valid in two dimensions. These results have been summarized in
a number of recent reviews\ref\rFV{{\it Dynamics of Fractal Surfaces},
edited by F. Family and T. Vicsek (World Scientific,
Singapore, 1991).}$^,$\ref\rKS{J.~Krug and H.~Spohn, in {\it Solids
Far From Equilibrium: Growth, Morphology and Defects}, edited by
 C.~Godreche (Cambridge University Press, Cambridge, 1991).}.

Let us now turn to the case of a line in three dimensions (Fig.\fFL ).
Fluctuations of the line can be indicated by a
a two dimensional vector $\br$. Even in an isotropic medium,
the drift velocity $\bv$ breaks the isotropy in \br\ by selecting
a direction. A gradient expansion up to second order
for the equation of motion
gives\ref\rEKlines{D.~Erta\c s and M.~Kardar,
Phys. Rev. Lett. {\bf 69}, 929 (1992).}
\eqn\eGrow{\eqalign{
\del_t r_\alpha=&\left[ K_1 \delta_{\alpha\beta}+
K_2 v_\alpha v_\beta\right] \del_x^2 r_\beta \cr
\noalign{\medskip}
&\quad+\left[ \lambda_1 (\delta_{\alpha\beta}v_\gamma
+\delta_{\alpha\gamma}v_\beta)
+\lambda_2 v_\alpha \delta_{\beta\gamma} + \lambda_3 v_\alpha
v_\beta v_\gamma
\right] {\del_x r_\beta \del_x r_\gamma \over 2}+ f_\alpha},}
with random force correlations
\eqn\eTT{\langle f_\alpha(x,t)f_\beta(x',t')\rangle=
2[T_1\delta_{\alpha\beta}+T_2v_\alpha v_\beta]\delta(x-x')\delta(t-t').}
Higher order nonlinearities can be similarly constructed but are in fact
irrelevant.
In terms of components parallel and perpendicular to the
velocity, the equations are
\eqn\eGro{\left\{\eqalign{\del_t \hl &= \Kl\del_x^2 \hl
+ {\lal  \over 2} (\del_x \hl)^2+ {\lalt \over 2}(\del_x \htr)^2
+\fl(x,t) \cr \noalign{\medskip}
\del_t\htr&=\Kt\del_x^2 \htr+\lat\del_x\hl\del_x \htr+\ft(x,t)} \right.
~~,}
with
\eqn\eD{\left\{\eqalign{\langle \fl(x,t)\fl(x',t')\rangle=&
2\Tl\delta(x-x')\delta(t-t') \cr \noalign{\medskip}
\langle \ft(x,t)\ft(x',t')\rangle=&
2\Tt\delta(x-x')\delta(t-t')}\right.~~.}
The noise-averaged correlations have a dynamic scaling form like
Eq.\escaling,
\eqn\eCor{\left\{\eqalign{
\langle[r_\parallel(x,t)-r_\parallel(x',t')]^2\rangle
&= |x-x'|^{2\zeta_\parallel}g_\parallel\left(
\frac{|t-t'|}{|x-x'|^{z_\parallel}}\right),\cr
\langle[r_\perp(x,t)-r_\perp(x',t')]^2\rangle
&= |x-x'|^{2\zeta_\perp}g_\perp\left(
\frac{|t-t'|}{|x-x'|^{z_\perp}}\right).\cr }\right.
}

\epsfysize=7cm\figure\fRGflow{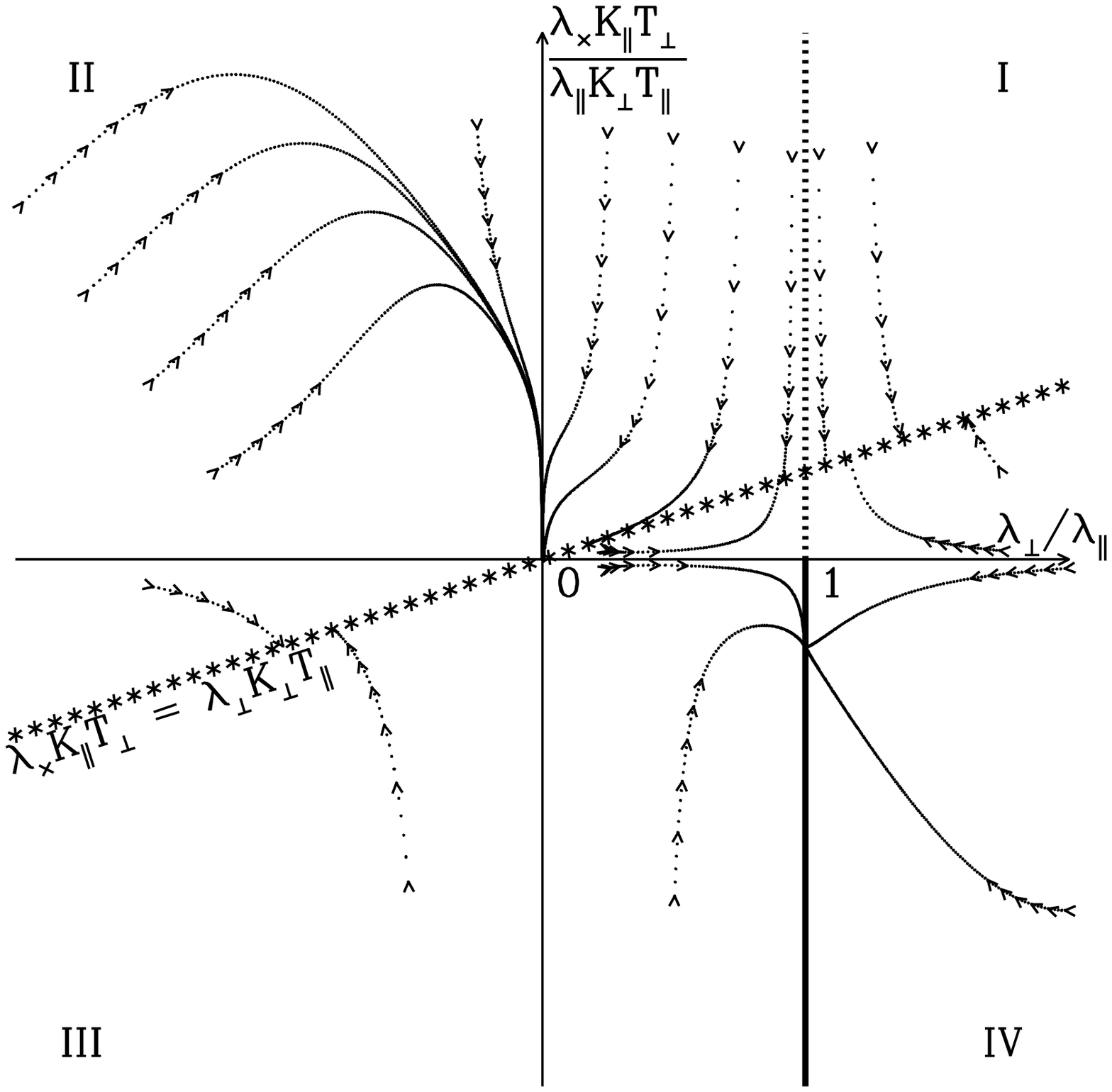}{A projection of RG flows
in the parameter space, for $n=1$ transverse components.}

In the absence of nonlinearities $(\lal=\lalt=\lat=0)$, Eqs.\eGro\
can easily be solved to give
$\zetal=\zetat=1/2$ and $\zl=\zt=2$. Simple dimensional counting
indicates that all three nonlinear terms are relevant
and may modify the exponents in Eq.\eCor.
Studies of related stochastic equations\ref\rTH{T.~Hwa, Phys. Rev. Lett.
{\bf 69}, 1552 (1992).}$^,$\ref
\rDW{D. Wolf, Phys. Rev. Lett. {\bf 67}, 1783 (1991).}\
indicate that interesting dynamic phase diagrams may emerge from
the competition between nonlinearities.
Let us assume that $\lal$ is positive and finite (its sign can be
changed by $\hl\to -\hl$), and focus on the dependence
of the scaling exponents on the ratios $\lat/\lal$ and $\lalt/\lal$, as
depicted in Fig.\fRGflow. (It is more convenient to set
the vertical axis to $\lalt\Kl\Tt/\lal\Kt\Tl$.)

The properties discussed for the KPZ equation can be extended to this
higher dimensional case:

\noindent{\bf 1.} {\it Galilean Invariance (GI):} Consider the infinitesimal
reparametrization
\eqn\eGal{\left\{\eqalign{x'=x+\lal\epsilon t ~&,~t'=t~,\cr
\hl'=\hl+\epsilon x ~&,~\htr'=\htr~.}\right.}
Eqs.\eGro\ are invariant under this transformation
provided that $\lal=\lat$.
Thus {\it along this line} in Fig.\fRGflow\ there is GI, which
once more implies the exponent identity
\eqn\eGInew{\zetal+\zl=2.}

\noindent{\bf 2.} {\it Fluctuation--Dissipation (FD) Condition:}
The Fokker--Planck
equation for the evolution of the joint probability
${\cal P}\left[ \hl(x),\htr(x)\right]$ has a stationary solution
\eqn\eSS{{\cal P}_0\propto\exp\left(-\int dx\left[{\Kl\over 2\Tl}(\del_x\hl)^2
+{\Kt\over 2\Tt}(\del_x\htr)^2\right]\right),}
provided that $\lalt\Kl\Tt=\lat\Kt\Tl$. Thus for this special choice of
parameters, depicted by a starred line in Fig.\fRGflow, if ${\cal P}$
converges to this solution, the long--time behavior of the correlation
functions in Eq.\eCor\ can be directly read off Eq.\eSS,
giving $\zetal=\zetat=1/2$.

\noindent{\bf 3.} {\it The Cole--Hopf (CH) Transformation} is an important
method for the exact study of solutions of the one component
nonlinear diffusion
equation\rJMB. Here we generalize this transformation to the complex
plane by defining, {\it for }$\lalt<0$,
\eqn\eCH{
\Psi(x,t)=\exp\left(\frac{\lal \hl(x,t)+i\sqrt{-\lal\lalt}\htr(x,t)}{2K}
\right).}
The linear diffusion equation
$$\del_t \Psi=K\del_x^2 \Psi +\mu(x,t)\Psi,$$

\noindent then leads to Eqs.\eGro\ if $\Kl=\Kt=K$ and $\lal=\lat$.
[Here ${\rm Re}(\mu)=\lal\fl/2K$ and ${\rm Im}(\mu)=
\sqrt{-\lal\lalt}\ft/2K$.]
This transformation enables an exact solution of the
{\it deterministic} equation, and further allows us to write the solution
to the {\it stochastic} equation in the form of a path integral
\eqn\ePath{\Psi(x,t)=\int_{(0,0)}^{(x,t)}{\cal D}x(\tau)
\exp  \left\{ -\int_0^t d\tau  \left[{{\dot x}^2 \over 2K}
+\mu(x,\tau)\right]  \right\} .}
Eq.\ePath\ has been extensively studied in connection with quantum
tunneling in a disordered medium\ref
\rMKSW{E. Medina, M. Kardar, Y. Shapir, and X.-R. Wang, Phys. Rev. Lett.
{\bf 62}, 941 (1989); E. Medina and M. Kardar, Phys. Rev. B {\bf 46}, 9984
(1992).},
with $\Psi$ representing the wave function. In particular, results for the
tunneling probability $|\Psi|^2$ suggest $\zl=3/2$ and
$\zetal=1/2$. The transverse fluctuations correspond to the phase
in the quantum problem which is not an observable. Hence this mapping does
not provide any information on $\zetat$ and $\zt$ which
are in fact observable for the moving line.

At the point $\lat=\lalt=0$, $\hl$ and $\htr$ decouple,
and $\zt=2$ while $\zl=3/2$. However, in general
$\zl=\zt=z$ unless the effective $\lat$ is zero. For example
at the intersection of the subspaces with GI and FD the
exponents $\zl=\zt=3/2$ are obtained from the
exponent identities. Dynamic RG recursion relations
can be computed to one--loop order\rEKlines$^,$\ref\rEKpoly{D.~Erta\c s and
M.~Kardar, Phys. Rev. E {\bf 48}, 1228 (1993).},
by standard methods of momentum-shell dynamic RG\rFNS$^,$\rMHKZ.

The renormalization of the seven parameters in Eqs.\eGro, generalized
to $n$ transverse directions, give the recursion relations
\eqn\eRR{\eqalign{
{d K_\parallel\over d\ell} &=
 K_\parallel\left[z-2+{1\over\pi}{\lambda_\parallel^2T_\parallel\over
4 K_\parallel^3}
+n{1\over\pi}{\lambda_\perp\lambda_\times T_\perp\over 4 K_\parallel
K_\perp^2}\right],\cr
{d K_\perp \over d\ell} &=
 K_\perp\left[z-2+{1\over\pi}{\lambda_\perp\big( (\lambda_\times
T_\perp/ K_\perp)
+(\lambda_\perp T_\parallel/ K_\parallel)\big) \over 2 K_\perp
( K_\perp+ K_\parallel)} \right. \cr
 & \hskip 1in \left. +{1\over\pi}{ K_\perp- K_\parallel \over  K_\perp+
K_\parallel}
{\lambda_\perp\big( (\lambda_\times T_\perp/ K_\perp)
-(\lambda_\perp T_\parallel/ K_\parallel)\big) \over  K_\perp
( K_\perp+ K_\parallel)}\right],\cr
{dT_\parallel \over d\ell} &=  T_\parallel\left[z-2\zeta_\parallel-1
+{1\over\pi}{\lambda_\parallel^2T_\parallel \over 4 K_\parallel^3}
\right]+n{1\over\pi}{\lambda_\times^2T_\perp^2 \over 4 K_\perp^3},\cr
{dT_\perp \over d\ell} &= T_\perp\left[z-2\zeta_\perp-1
+{1\over\pi}{\lambda_\perp^2T_\parallel \over  K_\perp K_\parallel
( K_\perp+ K_\parallel)}\right],\cr
{d\lambda_\parallel \over d\ell}  &= \lambda_\parallel\left[
\zeta_\parallel+z-2\right],\cr
{d\lambda_\perp \over d\ell}  &= \lambda_\perp\left[
\zeta_\parallel+z-2-{1\over\pi}
{\lambda_\parallel-\lambda_\perp \over ( K_\perp+ K_\parallel)^2}
{\big( (\lambda_\times T_\perp/ K_\perp)-(\lambda_\perp
T_\parallel/ K_\parallel)\big)}\right],\cr
{d\lambda_\times \over d\ell} &= \lambda_\times\left[
2\zeta_\perp-\zeta_\parallel+z-2
+{1\over\pi}{\lambda_\parallel K_\perp-\lambda_\perp K_\parallel
\over  K_\perp K_\parallel( K_\perp+ K_\parallel)}
{\big( (\lambda_\times T_\perp/ K_\perp)-(\lambda_\perp
T_\parallel/ K_\parallel)\big)}\right]. \cr
}}

The projections of the RG flows on the two parameter subspace shown in
Fig.\fRGflow\ are indicated by trajectories.
They naturally satisfy the constraints
imposed by the non--perturbative results: the subspace of GI is closed under
RG, while the FD condition appears as a {\it fixed line}. The RG
flows, and the corresponding exponents, are different in each quadrant
of Fig.\fRGflow, which implies that the scaling behavior is determined
by the relative signs of the three nonlinearities. This was confirmed by
numerical integrations\rEKlines$^,$\rEKpoly\ of
Eqs.\eGro,  performed for different sets of parameters. A summary of the
computed exponents are given in Table I.

The analysis of analytical and numerical results can be summarized
as follows:

$\lat\lalt>0$ : In this region, the scaling behavior is understood best.
The RG flows terminate on the fixed line where FD conditions
apply, hence $\zetal=\zetat=1/2$. All along this line, the one
loop RG exponent is $z=3/2$. These results are consistent with the numerical
simulations. The measured exponents rapidly converge to these values, except
when $\lat$ or $\lalt$ are small.

$\lalt=0$: In this case the equation for $\hl$ is the KPZ equation \eKPZ,
thus $\zetal=1/2$ and $\zl=3/2$. The fluctuations in
$\hl$ act as a strong (multiplicative and correlated) noise on $\htr$.
The one--loop RG yields the exponents
$\zt=3/2,\ \zetat=0.75$\ for $\lat>0$, while
a negative $\lat$ scales to 0 suggesting $z_\perp>z_\para$.
Simulations are consistent with the RG calculations for $\lat>0$, yielding
$\zeta_\perp=0.72$, surprisingly close to the one--loop RG value.
For $\lat<0$, simulations indicate $z_\perp\approx 2$ and
$\zeta_\perp\approx 2/3$ along with the expected values for the
longitudinal exponents.

$\lat=0$: The transverse fluctuations satisfy a simple diffusion equation
with $\zetat=1/2$ and $\zt=2$. Through the term $\lalt(\del_x \htr)^2/2$,
these fluctuations act as a correlated noise\rMHKZ\ for the longitudinal
mode. A naive application of the results of this  reference\rMHKZ\  give
$\zetal=2/3$ and $\zl=4/3$.
Quite surprisingly, simulations indicate different behavior depending on
the sign of $\lalt$.
For $\lalt<0$, $\zl\approx 3/2$ and $\zetal\approx1/2$
whereas for $\lalt>0$, longitudinal fluctuations are much stronger,
resulting in $\zl\approx 1.18$ and $\zetal\approx 0.84$. Actually,
$\zetal$ increases steadily with system size, suggesting a breakdown
of dynamic scaling, due to a change of sign in $\lat\lalt$.
This dependence on the sign of $\lalt$ may reflect
the fundamental difference between behavior
in quadrants II and IV of Fig.\fRGflow.

\epsfysize=10cm
\medskip
\centerline{\epsffile{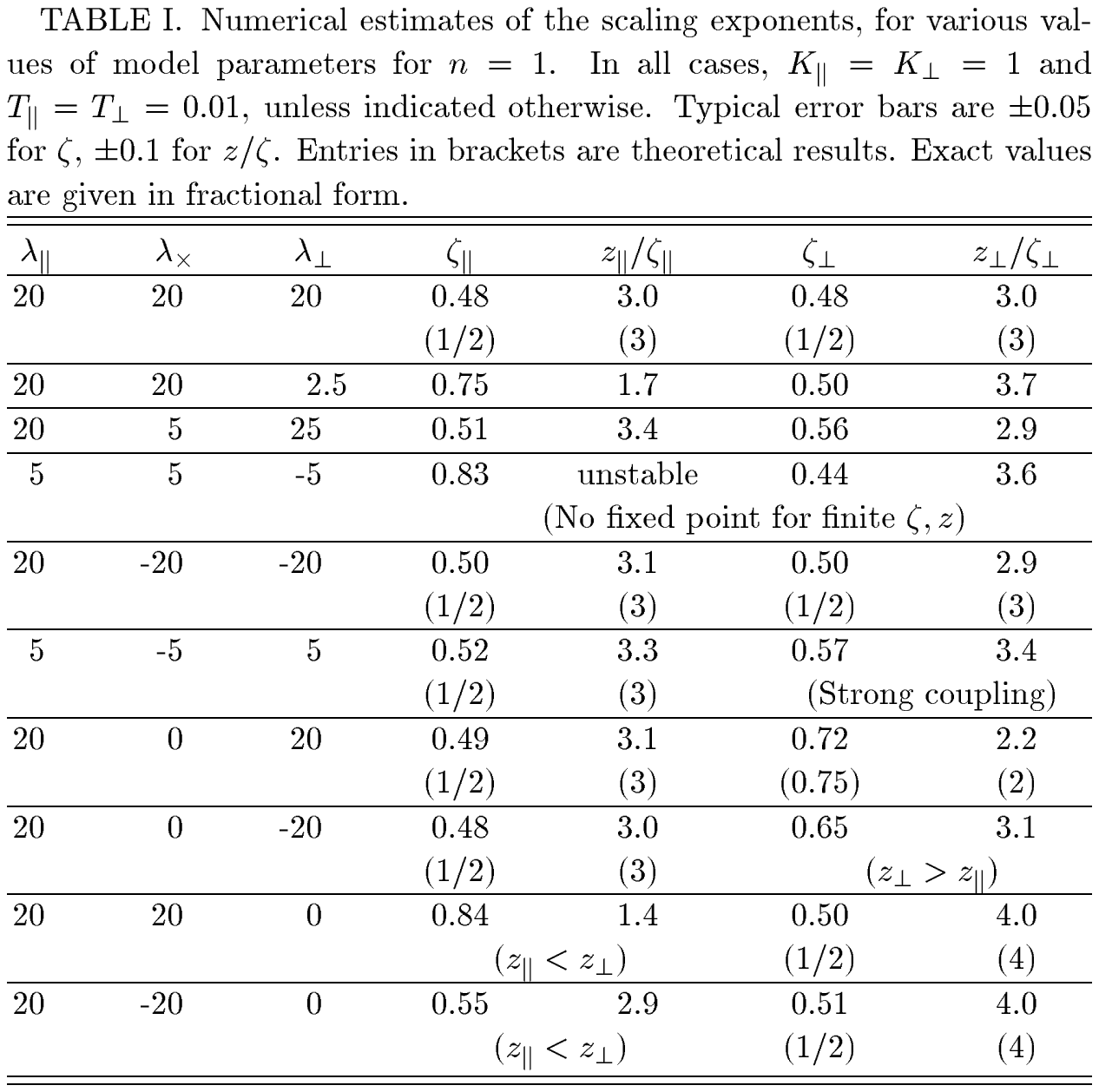}}
\medskip

$\lat<0$ {\it and} $\lalt>0$: The analysis of this region (II) is the most
difficult in that the RG flows do not converge upon a finite fixed point
{\it and } $\lat\to 0$, which may signal the breakdown of dynamic scaling.
Simulations indicate strong longitudinal fluctuations that lead
to instabilities in the discrete integration scheme, excluding the
possibility of measuring the exponents reliably.

$\lat>0$ {\it and} $\lalt<0$: The {\it projected\/} RG flows in this
quadrant (IV) converge to the point $\lat/\lal=1$ and $\lalt\Tt\Kl/
\lal\Tl\Kt=-1$. This is actually not a fixed point, as
$K_\para$ and $K_\perp$ scale to infinity. The applicability
of the CH transformation to this point implies $\zl=3/2$ and $\zetal=1/2$.
Since $\lat$ is finite, $\zt=\zl=3/2$ is expected, but this does
not give any information on $\zetat$. Simulations indicate strong
transverse fluctuations and suffer from difficulties similar to
those in region II.

Eqs.\eGro\ are the simplest nonlinear, local,
and dissipative equations that govern the fluctuations of a moving
line in a random medium. They can be easily generalized to describe
the time evolution of a manifold with arbitrary internal
($\bx \in R^{d}$) and external
($\br \in R^{n+1}$) dimensions, and to the motion of curves
that are not necessarily stretched in a particular direction.
Since the derivation only involves
general symmetry arguments, the given results are widely applicable
to a number of seemingly unrelated systems.
We will discuss one application to drifting polymers in more
detail in the next lecture, explicitly demonstrating the origin
of the nonlinear terms starting from more fundamental
hydrodynamic equations. A simple model of crack front propagation
in three dimensions\ref\rBBLP{J.~P.~Bouchaud, E.~Bouchaud, G.~Lapasset,
and J.~Planes, preprint (1993).}\ also arrives at Eqs.\eGro, implying the
self-affine
structure of the crack surface after the front has passed.
\medskip
%% LECTURE 4
\vskip .15truein \noindent{\bf IV. NONLINEAR RELAXATION OF DRIFTING
POLYMERS}\smallskip\par
The dynamics of polymers in fluids is of much theoretical interest
and has been extensively studied\ref\rDoi{M.~Doi and S.F.~Edwards,
{\it Theory of Polymer Dynamics},
Oxford University Press (1986).}$^,$\ref\rDeG{P.G.~de~Gennes,
{\it Scaling Concepts in Polymer Physics},
Cornell University Press (1979).}. The combination
of polymer flexibility, interactions, and hydrodynamics make a
first principles approach to the problem quite difficult. There are,
however, a number of phenomenological studies that describe various
aspects of this problem\ref\rBrd{R.B.~Bird, {\it Dynamics of Polymeric
Physics},
Vols. 1-2,
Wiley, New York (1987).}.
\epsfysize=7cm\figure\fPoly{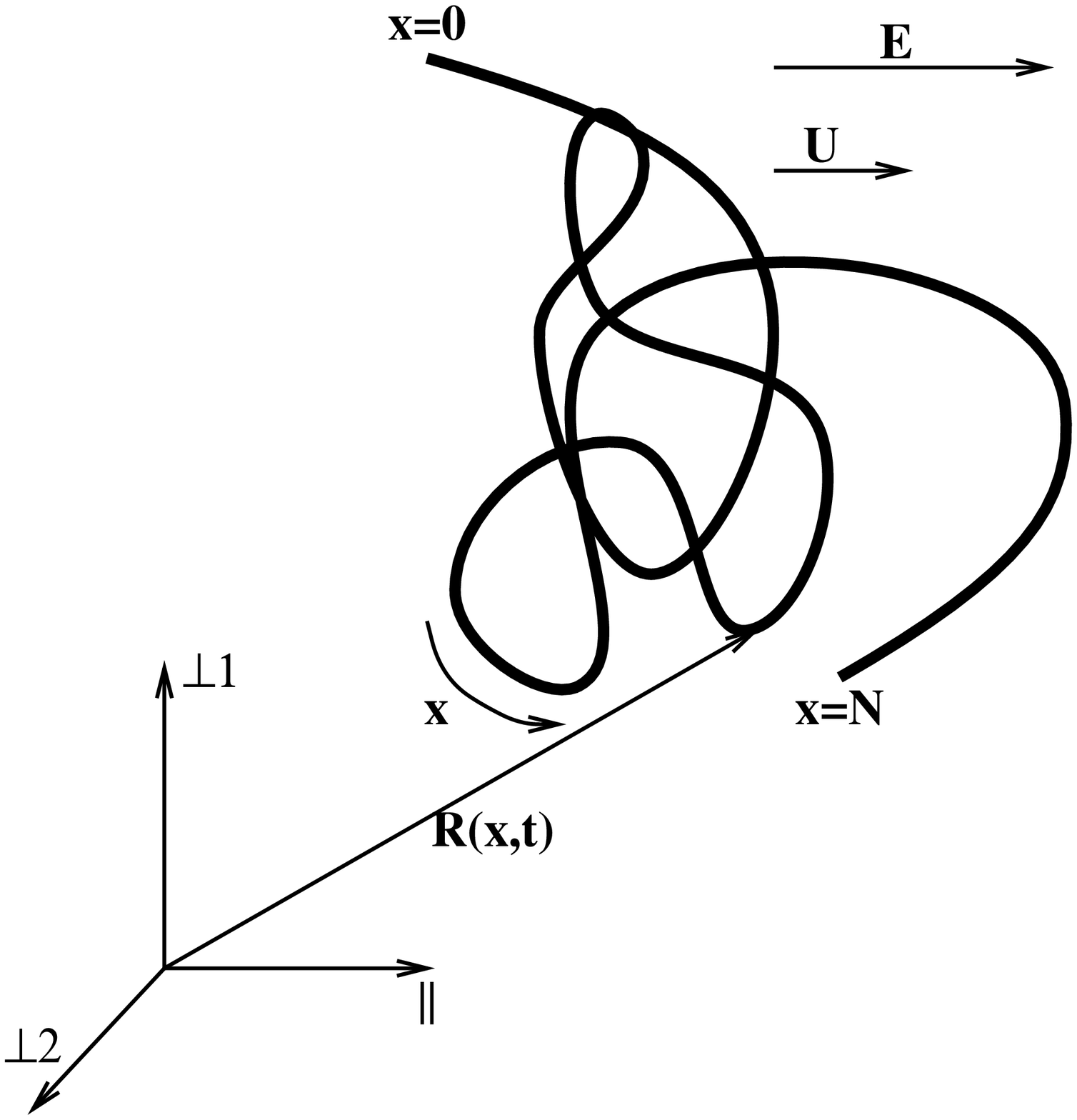}{The configuration of a polymer.}
One of the simplest is the
Rouse model\ref\rRs{P.E.~Rouse, J. Chem. Phys. {\bf 21}, 1272
(1953).}: The configuration of the polymer at
time $t$ is described by a vector ${\bf R}(x,t)$, where $x \in [0,N]$
is a continuous variable replacing the discrete monomer index (see Fig.\fPoly).

Ignoring inertial effects, the relaxation of the polymer in a viscous
medium is approximated by
\eqn\Rouseeq{
\partial_t{\bf R}(x,t)=\mu{\bf F(R}(x,t){\bf )} =
K\partial_x^2{\bf R}(x,t)+{\eta}(x,t),
}
where $\mu$ is the mobility. The force {\bf F} has a
contribution
from interactions with near neighbors that are treated as
springs.
Steric and other interactions are ignored. The effect of the
medium
is represented by the random forces $\eta$ with zero
mean.
The Rouse model is a linear Langevin equation that is easily
solved.
It predicts that the mean square radius of gyration,
$R_g^2=\langle
|{\bf R}-\langle {\bf R}\rangle |^2\rangle $, is proportional
to the
polymer size $N$, and the largest relaxation times scale as the
fourth power of the wave number, (i.e., in dynamic light
scattering
experiments, the half width at half maximum of the scattering
amplitude scales as the fourth power of the scattering
wave vector
{\bf q}). These results can be summarized as $R_g\sim N^\nu $
and
$\Gamma({\bf q}) \sim q^z$, where $\nu$ and $z$ are called the
{\it swelling} and {\it dynamic} exponents, respectively\ref
\rFoot{We have changed the notation to confer with the
traditions of polymer science. $\nu$ is
$\zeta$ and $z$ is $z/\zeta$ in terms of the notation used previously.}.
Thus, for the Rouse Model, $\nu=1/2$ and $z=4$.

The Rouse model ignores hydrodynamic interactions mediated by
the
fluid.  These effects were originally considered by Kirkwood
and
Risemann\ref\rRis{J.~Kirkwood and J.~Risemann, J. Chem. Phys.
{\bf 16}, 565 (1948).}\ and later on by Zimm\ref\rZm{B.H.~Zimm,
J. Chem. Phys. {\bf 24}, 269 (1956).}. The basic
idea is that the motion of each monomer modifies the flow
field
at large distances. Consequently, each monomer experiences an
additional velocity
\eqn\eZimm{
\delta_H\partial_t{\bf R}(x,t) = {1\over{8\pi\eta_s}}\int
dx'{{{\bf F}(x')r_{xx'}^2+({\bf F}(x')\cdot{\bf r}_{xx'})
{\bf r}_{xx'}}\over {|{\bf r}_{xx'}|^3}} \approx \int dx'
{\gamma\over{|x-x'|^\nu}}\partial_x^2{\bf R},
}
where ${\bf r}_{xx'}={\bf R}(x)-{\bf R}(x')$ and the final
approximation is obtained by replacing the actual distance
between
two monomers by their average value.  The modified equation is
still linear in ${\bf R}$ and easily solved.  The main result
is the speeding up of the relaxation dynamics as the exponent
$z$ changes from 4 to 3.  Most experiments on polymer
dynamics\ref\rAdam{See, for example, M.~Adam and M.~Delsanti,
Macromolecules {\bf 10}, 1229 (1977).}\
indeed measure exponents close to 3. Rouse
dynamics is still important in other circumstances, such as
diffusion of a polymer in a solid matrix, stress and
viscoelasticity in concentrated polymer solutions, and is also
applicable to relaxation times in Monte Carlo simulations.

Since both of these models are linear, the dynamics remains
invariant in the center of mass coordinates upon the
application
of a uniform external force. Hence the results for a drifting
polymer are identical to a stationary one.  This conclusion is
in fact not correct due to the hydrodynamic interactions.
For example, consider a rodlike
conformation of the polymer with monomer length $b_0$ where
$\partial_xR_\alpha=b_0t_\alpha$ everywhere on the polymer, so that the
elastic (Rouse) force vanishes. If a uniform force ${\bf E}$ per
monomer acts on this rod, the velocity of the rod can be solved using
Kirkwood Theory, and the result is\rDoi
\eqn\erod{
{\bf v} = {(-\ln{\kappa}) \over 4\pi\eta_s b_0}{\bf E}\cdot
\left[{\bf I} + {\bf t}{\bf t}\right].}
In the above equation, $\eta_s$ is the solvent viscosity, ${\bf t}$
 is the unit tangent vector,  $\kappa= 2b/b_0N$ is the ratio
 of the  width $b$  to the half length $b_0N/2$ of the polymer.
A more detailed calculation
of the velocity in the more general case of an arbitrarily shaped
slender body by Khayat and Cox\ref\rKhayat{R.E.~Khayat and R.G.~Cox,
J. Fluid. Mech. {\bf 209}, 435 (1989).}\ shows that
{\it nonlocal} contributions to the hydrodynamic force, which depend
on the whole shape of the polymer rather than the local orientation,
are ${\cal O}(1/(\ln{\kappa})^2)$. Therefore, corrections to
Eq.\erod\ are small when $N \gg b/b_0$.

Incorporating this tilt dependence of polymer mobility requires adding
terms nonlinear in the tilt, $\partial_x {\bf r}$, to a {\it local} equation of
motion. Since the overall force (or velocity) is the only vector breaking
the isotropy of the fluid, the structure of these nonlinear terms must be
identical to eq.\eGrow. Thus in terms of the fluctuations parallel and
perpendicular to the average drift, we again recover the equations,
\eqn\eKPZE{\left\{\eqalign{
\partial_tR_\parallel &= U_\parallel +
K_\parallel\partial_x^2R_\parallel +
{{\lambda_\parallel}\over 2}(\partial_xR_\parallel)^2 +
{{\lambda_\times}\over 2}\sum_{i=1}^2(\partial_xR_{\perp i})^2
+
\eta_\parallel(x,t), \cr
\partial_tR_{\perp i} &= K_\perp\partial_x^2R_{\perp i} +
\lambda_\perp\partial_xR_\parallel\partial_xR_{\perp i} +
\eta_{\perp i}(x,t), \cr}\right.
}
where $\{\perp\!i\}$ refers to the 2 transverse
coordinates
of the monomer positions.  The noise is assumed to be white
and
gaussian but need not be isotropic, i.e.
\eqn\eNoise{\left\{\eqalign{
\langle\eta_\parallel(x,t)\eta_\parallel(x',t')\rangle &=
2T_\parallel\delta(x-x')\delta(t-t'), \cr
\langle\eta_{\perp i}(x,t)\eta_{\perp j}(x',t')\rangle &=
2T_\perp\delta_{i,j}\delta(x-x')\delta(t-t'). \cr}\right.
}
At zero average velocity, the system becomes isotropic and the equations of
motion must
coincide with the Rouse model.  Therefore,
$\{\lambda_\parallel,
\lambda_\times,\lambda_\perp,U,K_\parallel-K_\perp,T_\parallel
-T_\perp\}$ are all proportional to $E$ for small forces.
The relevance of these nonlinear terms are determined by
the dimensionless scaling variable
$$ y=\left({U\over U^*}\right)N^{1/2}, $$
where $U^*$ is a characteristic microscopic velocity
associated with
monomer motion and is roughly 10-20 m/s for polystyrene in
benzene. The variable $y$ is proportional to another
dimensionless parameter, the Reynolds number $Re$, which
determines the breakdown of hydrodynamic equations and
onset of turbulence. However, typically $Re \ll y$, and the
hydrodynamic equations are valid for moderately large $y$.
Eqs. \eKPZE\ describe the static
and dynamical scaling properties of the nonlinear and
anisotropic
regime when $U>U^*N^{-1/2}$.

Eq.\eKPZE\ is just a slight variation from \eGro, with two
transverse components instead of one. Thus, the results
discussed in the previous lecture apply. A more detailed
calculation of the nonlinear terms from hydrodynamics\ref
\rEKapp{See Appendices A and B of our longer paper\rEKpoly.}\
shows that all three nonlinearities are positive for small driving forces.
In this case, the asymptotic scaling exponents are isotropic,
with $\nu=1/2$ and $z=3$. However, the fixed points of the
RG transformation are in general anisotropic, which implies
a kinetically induced form birefringence {\it in the absence
of external velocity gradients}. This is in marked contrast
with standard theories of polymer dynamics where a uniform
driving force has essentially no effect on the internal modes
of the polymer.

When one of the nonlinearities approaches to zero, the swelling
exponents may become anisotropic and the polymer elongates or
compresses along the longitudinal direction.  However, the
experimental path in the parameter space as a function of $E$
is not
known and not all of the different scaling regimes correspond
to
actual physical situations.  The scaling results found by the
RG
analysis are verified by direct integration of equations, as
mentioned in the earlier lectures. A more detailed discussion
of the analysis and results can be found in our earlier
work\rEKpoly.

In constructing equations \eKPZE, we only allowed for local
effects, and ignored the nonlocalities that are the hallmark of
hydrodynamics. One consequence of hydrodynamic interactions
is the {\it back-flow} velocity in Eq.\eZimm\ that can be added
to the evolution equations \eKPZE. Dimensional analysis
gives the recursion relation
\eqn\eZimmRR{\frac{\del\gamma}{\del\ell}=\gamma\left[
\nu z -1-(d-2)\nu\right]+O(\gamma^2),}
which implies that, at the nonlinear fixed point, this
additional term is surprisingly irrelevant for $d>3$,
and $z=3$ due to the nonlinearities. For $d<3$, $z=d$ due to
hydrodynamics, and the nonlinear terms are irrelevant.
The situation in three dimensions is unclear,
but a change in the exponents is unlikely. Similarly, one
could consider the effect of self-avoidance by including
the force generated by a softly repulsive contact potential
\eqn\eContact{{b\over2}\int\,dx\,dx'\,{\cal V}
\left(\br(x)-\br(x')\right).}
The relevance of this term is also controlled by the scaling
dimension $y_b=\nu z -1-(d-2)\nu$, and therefore this effect is
marginal in three dimensions at the nonlinear fixed point,
in contrast with both Rouse and Zimm models where
self-avoidance becomes relevant below four dimensions.
Unfortunately, one is ultimately forced to consider
non-local {\it and} nonlinear terms based on similar
grounds, and such terms are indeed relevant below
four dimensions.
In some cases, local or global arclength conservation may be
an important consideration in writing down a dynamics for the
system. However, a
local description is likely to be more correct in a more
complicated system with screening effects (motion in a gel
that screens hydrodynamic interactions) where a first
principles approach becomes even more intractable. Therefore,
this model is an important starting point towards understanding
the scaling behavior of polymers under a uniform drift, a
problem with great technological importance.
\medskip
\vskip .15truein \noindent{\bf ACKNOWLEDGEMENTS}\smallskip\par

We have benefited from discussions with O.~Narayan and L.-H.~Tang.
This research was
supported by grants from the NSF (DMR-93-03667 and PYI/DMR-89-58061),
and the MIT/INTEVEP collaborative program.

\bigskip\immediate\closeout\rfile
\noindent{\bf REFERENCES}\bigskip
{\catcode`\@=11\escapechar=`  \input refs.tmp\vfill\eject}
\end